\documentclass[11pt]{article}
\usepackage[dvips]{graphicx}
\usepackage{amsfonts}
\usepackage{amssymb}
\usepackage{amsmath}
\usepackage{epsfig}
\usepackage{multirow}
\usepackage{latexsym}
\usepackage{graphicx}
\usepackage{float}
\usepackage{psfrag}

\usepackage[nosort]{cite}

\newcommand{\nn}{\nonumber}
\newcommand{\be}{\begin{equation}}
\newcommand{\ee}{\end{equation}}
\newcommand{\bea}{\begin{eqnarray}}
\newcommand{\eea}{\end{eqnarray}}

\newcommand{\lsim}{\lesssim}

\newcommand{\beqa}{\begin{eqnarray}}
\newcommand{\eeqa}{\end{eqnarray}}

\topmargin
-0.7cm
\textwidth
15.5cm
\textheight
22.8cm
\oddsidemargin
0.7cm
\evensidemargin
1.2cm

\begin{document}

\makeatletter
\@addtoreset{equation}{section}
\makeatother
\renewcommand{\theequation}{\thesection.\arabic{equation}}
\pagestyle{empty}
\vspace*{0.1cm}
\rightline{CPHT-RR034.0712}
\rightline{CERN-PH-TH/2012-183}
\vspace{2.2cm}
\begin{center}
\Large{\bf Universal contributions to scalar masses from five dimensional supergravity\\[12mm]}
\large{Emilian Dudas$^{1,2}$,\; Gero von Gersdorff$^2$ \\[5mm]}
\small{$^1$ CERN, PH-TH Division, CH-1211 Gen\`eve 23, Switzerland}\\
\vspace{0.1cm}
\small{$^2$ Centre de Physique Th\'eorique, \'Ecole Polytechnique, CNRS, 91128 Palaiseau, France.}\\
\vspace{0.1cm}

\small{E-mail: emilian.dudas@cpht.polytechnique.fr, gero.gersdorff@cpht.polytechnique.fr  } \\[12mm]
\small{\bf Abstract} \\[3mm]
\end{center}
\begin{center}
\begin{minipage}[h]{16.0cm}
We compute the effective Kahler potential for matter fields in warped compactifications, starting from five dimensional gauged supergravity, as a function of the matter fields localization. We show that truncation to zero modes is inconsistent and the tree-level exchange of
the massive gravitational multiplet is needed for consistency of the four-dimensional theory. In addition to the standard
Kahler coming from dimensional reduction, we find the quartic correction coming from integrating out the gravity multiplet.
We apply our result to the computation of scalar masses, by assuming that the SUSY breaking field is a bulk hypermultiplet.
In the limit of extreme opposite localization of the matter and the spurion fields, we find zero scalar masses, consistent with
sequestering arguments. Surprisingly enough, for all the other cases the scalar masses are tachyonic. This suggests the holographic
interpretation that a CFT sector always generates operators contributing in a tachyonic way to scalar masses.
Viability of warped supersymmetric compactifications necessarily asks then for additional contributions. We discuss
the case of additional bulk vector multiplets with mixed boundary conditions, which is a particularly simple and attractive way to generate large positive scalar masses. We show that in this case successful fermion mass matrices implies highly degenerate scalar masses
for the first two generations of squarks and sleptons.

\end{minipage}
\end{center}
\newpage
\setcounter{page}{1}
\pagestyle{plain}
\renewcommand{\thefootnote}{\arabic{footnote}}
\setcounter{footnote}{0}


\tableofcontents


\section{Introduction}
\label{sec:intro}
Soft supersymmetry breaking terms for visible sector fields have various contributions in a higher-dimensional
field or string theory. For example, at tree-level $F$-term contributions to scalar masses are given by
\be
m_{a \bar b}^2 \ = \ m_{3/2}^2 \delta_{a \bar b} - (K_{a \bar a} K_{b \bar b})^{-1/2} F^{\alpha}
R_{a {\bar b} \alpha {\bar \beta}} F^{\bar \beta} \ , \label{intro1}
\ee
where $K_{a \bar a}$ is the wave function of the matter field $\Phi_a$, $F^{\alpha}$ is the auxiliary field of the SUSY breaking
field $\Phi_{\alpha}$ and $R_{a {\bar b} \alpha {\bar \beta}}$ the Riemann curvature encoding the quartic couplings between the matter
fields and the SUSY breaking ones. The first term in (\ref{intro1}) is the famous universal (mSUGRA) contribution, whereas the
second term depends on the detailed form of the Kahler potential and can have various origins. It can have a gravitational origin as typical in higher-dimensional supergravity/superstrings compactifications or it can be induced at lower scales by field theory dynamics, like in gauge mediation.

In this paper we would like to comment on a further possibility for such contributions. The class of models they can arise in are those with a warped fifth dimension with a Kaluza Klein (KK) scale much smaller than the Planck scale. Via the gauge-gravity correspondence these models are dual to
strongly coupled (large-$N_c$) gauge theories with meson resonances described by the KK modes. Models of this kind have extensively been studied in phenomenological applications, since they can provide a theory of flavor \cite{flavor, tony}, induce supersymmetry breaking \cite{Gherghetta:2000kr} or even alleviate the little hierarchy problem  in the extreme case of a TeV KK scale \cite{little}.

The first point that we expand in detail in the present paper, sometimes overlooked in the literature, is that a simple
truncation of the higher-dimensional action is not enough for finding the consistent 4d two-derivative Lagrangian.
One could expect that simply integrating over the 5d Lagrangian would
yield a supersymmetric 4d effective Lagrangian, and contributions from KK exchange would yield a -- equally supersymmetric -- correction to it.
As we will see, this naive expectation is incorrect, and the two contributions have to be combined to give a supersymmetric result.
We show that the contributions from the entire KK tower can be computed in closed form,
 and use this to calculate explicitly the quartic corrections to the Kahler potential in 4d as a function of the localization
of bulk fields. We then apply the result to the calculation of soft masses for matter fields, under the assumption that matter and
the spurion originate from zero modes of bulk hypermultiplets. Our result is consistent with the vanishing of scalar masses for the sequestered case. However, surprisingly we find that for all other localization patterns the scalar masses are tachyonic\footnote{After
completing our work, we learned that this result was previously obtained, for the case of symplectic hypermultiplet spaces, by a different method in \cite{sakamura,abe}. We thank Y. Sakamura and T. Higaki for pointing out these references to us.}. In order to cure this problem, we study a simple generalization which includes
abelian bulk vector multiplets with $(+,-)$ and $(-,+)$ boundary conditions. We show that the KK abelian gauge fields exchange generate
generically larger than gravity contributions, the sign of which is determined by the abelian charges. We also point out that these
contributions are independent of localization and therefore universal (for equal abelian charges) for a large region of the parameter space. Extra-dimensional localization generates successful Yukawa hierarchies, compatible and actually implying degeneracy of the first two generations of squarks and leptons. We view therefore this possibility as a viable solution to the supersymmetric flavor problem.

The examples we study here are  $AdS_5$ curved compactifications, which are supersymmetric generalizations of the
RS type compactifications \cite{rs1} with bulk fields localized by profiles of their wavefunctions \cite{tony}.
These models are conjectured to have an AdS/CFT 4d holographic interpretation, the 4d fields being elementary if localized on the
UV brane and composite if localized on the IR brane. Moreover, the 5d supergravity fields are dual to the Ferrara-Zumino (FZ) supercurrent \cite{Ferrara:1974pz} containing the $R$ current and the energy-momentum tensor of the gauge theory. Once the theory confines these currents are expected to excite a discrete spectrum of mesons whose masses and interactions are described dually by the KK Lagrangian of 5d supergravity. In particular their masses are given by the scale related to the position of the infrared brane (i.e., the warped-down Planck scale). The CFT origin of the new contributions to the soft terms is thus an effect of the heavy mesons (of the FZ current) in the low-energy (confined) phase of the gauge theory. As it turns out, the quartic Kahler operators we find only depend on the $R$ charges of the matter fields and the IR scale of the theory.

The paper is organized as follows. In Sec.~\ref{sec:gauged} we review the couplings of 5d supergravity to matter as far as they will be needed for writing the low energy matter Lagrangian.
In Sec.~\ref{sec:kk} we first derive the general low energy Lagrangian obtained form integrating out the KK modes of the supergravity multiplet. We then focus on four-fermion terms and reconstruct from them the quartic terms in the effective Kahler potential. In Sec.~\ref{sec:scalar} we derive the soft scalar masses obtained from this Kahler potential under the assumption that the susy breaking spurion is the zero mode of a bulk hypermultiplet. In Sec.~\ref{sec:vector} we discuss the details of the bulk abelian vector
multiplets exchange, the resulting generated scalar and other soft masses and a brief discussion of the low-energy phenomenology. Finally in Sec.~\ref{sec:summary} we present our conclusions and comment on some open questions. Several more technical issues are relegated to appendices. In App.~\ref{wfprop} we compute the propagators of the bosonic supergravity fields in the $AdS_5$ background.
In App.~\ref{zeromomentum} we give some details on the evaluation of the effective low-energy action skipped in Sec~\ref{sec:kk}. Finally, in App.~\ref{app:scalars} we explicitly compute the contributions to the effective action of the scalars (the two-derivative action at quartic order in the fields) and show that it matches the fermionic result in Sec.~\ref{sec:kk}, as expected from supersymmetry.

\section{The setup: five dimensional gauged supergravity}
\label{sec:gauged}

The five dimensional supergravity Lagrangian with and without matter has been developed by many authors \cite{sugralit,Kugo:2000af}. Here we will rely mostly on Refs.~\cite{Kugo:2000af} who in particular consider the presence of boundaries.
Since we will require an $AdS_5$ vacuum we will need to gauge a $U(1)_R$ subgroup of the $SU(2)_R$ automorphism group under which the two supercharges transform.

The most general couplings of hypermultiplets to supergravity are described by special non-linear sigma models known as quaternionic Kahler manifolds. In contrast to $N=1$ supergravity in 4d, these sigma models are tightly constrained, in particular their curvature is determined by supersymmetry, which in turn fixes the self coupling of the sigma model and the four-fermion interactions of the matter fields.

Instead of writing the full supergravity Lagrangian, we will restrict ourselves to the couplings that are relevant for our purpose. The interested reader can find the complete Lagrangians in Refs.~\cite{Kugo:2000af} from which all the interactions below can be extracted straightforwardly. For definiteness we will also only consider two classes of sigma models, the spaces $\frac{USp(2,2n)}{USp(2) \times USp(2n)}$ and $\frac{U(2,n)}{U(2) \times U(n)}$. They are the simplest ones in the sense that they can be constructed with the smallest amount of compensator hypermultiplets (one in the case of the symplectic coset, two for the unitary case). Both of these manifolds have (real) dimension $4n$ (i.e.~they both describe $n$ physical hypermultiplets), are non-compact and have negative curvature. The unitary space with $n=1$ is known to describe the universal hypermultiplet arising from string compactifications.

The field content of 5d $N=2$ supergravity consists of the metric $(h_{MN})$ containing 5 degrees of freedom, the graviphoton $(A_M)$ containing 3 degrees of freedom, and the gravitino ($\Psi_M^i$) with 8 degrees of freedom.
The graviphoton coupling constant $g_R$ is an a priori free parameter. The gauging will give rise to a negative cosmological constant term $-16M^6g_R^2$, leading to an $AdS_5$ vacuum with curvature $k$ related to $g_R$ as
\be
g_R^2=\frac{3\,k^2}{4\,M^3}\,.
\label{Rcoupling}
\ee

In the following we will only keep terms in the matter Lagrangian that contribute to the sources of the bosonic supergravity fields as well as the self-interactions of the scalars and fermions.
Let us split the matter-gravity interactions into two parts
\be
\mathcal L^{\rm matter}=\mathcal L^{\rm fermion}+\mathcal L^{\rm scalar}+\dots
\ee
where the ellipsis denotes terms containing both fermions and scalars that are also not relevant for the present work.
Turning to the fermions first, the relevant interactions are
\footnote{Our conventions are as follows. $A,B,\dots$ ($a,b\dots$) denote 5d (4d) tangent space indices $M,N,\dots$ ($\mu,\nu\dots$) coordinate indices, the metric is mostly plus, $\eta_{AB}=(-1,1,1,1,1)$ and the convention for the gamma matrices is
\be\gamma^{\mu} \ = \  -i \ \begin{pmatrix} 0 & \sigma^{\mu} \\ {\bar \sigma}^{\mu} & 0
\end{pmatrix} \quad , \quad  \gamma^{5} = \begin{pmatrix} 1 & 0 \\ 0 & -1 \end{pmatrix}\, , \ee
where $\sigma^{\mu} = (1,\sigma^i)$, ${\bar \sigma}^{\mu} = (1,-\sigma^i)$. Antisymmetrization of the gamma matrices is with strength one. The Einstein-Hilbert term is normalized as $\mathcal L^{\rm EH}=-M^3R$.
}
\be
\mathcal L^{\rm fermion}
= -i\, \bar\Psi\,\gamma^A\,D_A\,\Psi-i\,m_\Psi\,\bar\Psi\,\Psi\,
-\sqrt{\frac{3}{64\,M^3}}\, \bar\Psi\,\gamma^{AB}\,\Psi\,F_{AB}
+\mathcal L^{4f}
\label{Lfermion}
\ee
where $\Psi=(\psi_L,\psi_R)$ denotes 5d Dirac fermions (we have explicitly solved for all reality constraints).
Notice the dipole interaction with the graviphoton field strength $F_{AB}$.
The covariant derivatives are given by $D_M=\partial_M+\Gamma_M+ig_R\,q_R\,A_M$.
The gauging also relates the mass term and 5d $R$ charge as
\be
m_{\Psi_i}=c_i\,k\,,\qquad q_{\Psi_i}^R=-\frac{2}{3}\,c_i\,.
\label{masscharge}
\ee
The four-fermion terms are slightly different for the symplectic and unitary cosets:\footnote{These 4-fermion terms arise from integrating out auxiliary fields of the 5d supergravity. In particular, in the unitary coset model there is an additional non-propagating $U(1)$ gauge field \cite{Kugo:2000af}, giving rise to the second term in Eq.~(\ref{4fU}).}
\bea
\mathcal L^{4f}_S&=&-\frac{1}{64\,M^3} (\bar\Psi\,\gamma^{AB}\,\Psi)^2\,,\label{4fS}\\
\mathcal L^{4f}_U&=&-\frac{1}{64\,M^3} (\bar\Psi\,\gamma^{AB}\,\Psi)^2
					-\frac{1}{16\,M^3} (\bar\Psi\,\gamma^{A}\,\Psi)^2 \,.\label{4fU}
\eea

In the scalar sector we can write
\be
\mathcal L^{\rm scalar}=
\mathcal L^{\rm  \Sigma}-V(\Phi)
\label{Lscalar}
\ee
The sigma model Lagrangians in the two cases are as follows. In the unitary coset, it can be obtained from a Kahler potential
\be
\mathcal L^\Sigma_U\ =\ -\, \partial_{\bar\Phi_i}\partial^{\ }_{\Phi_J}K_{U}\ D_M\,\bar\Phi_I\, D_M\,\Phi_J
\ee
with the Kahler potential given by \cite{Kugo:2000af} ($\kappa^{-1}=2\,M^3$)
\be
K_U\ =\ -\kappa^{-1}\,\ln\left[(1-\kappa\,|\Phi_1|^2)(1-\kappa\,|\Phi_2|^2)-\kappa^2\,|\Phi_2^\dagger\Phi^{\ }_1|^2\right]
\label{kahleruni}
\ee
In the symplectic coset, the sigma model Lagrangian cannot be obtained from a Kahler potential \cite{deWit:1983rz}. We thus give its explicit form:
\bea
\mathcal L_S^{\Sigma}&=&-F(\Phi)^{-1}\biggl(
		|D_M\,\Phi_1|^2+|D_M\,\Phi_2|^2\biggr)\nn\\
		&&+\frac{\kappa}{2}\,
		F(\Phi)^{-2}\,\biggl(\left|\Phi_1^\dagger\,D_M\Phi_1+\Phi_2^\dagger\,D_M\Phi_2\right|^2
				-\left|\Phi_2^\dagger\, D_M\Phi_1-D_M\Phi_2^\dagger\,\Phi_1\right|^2
		\biggr) \label{kahlersym}
\eea
with $F(\Phi)=1-\frac{\kappa}{2}(|\Phi_1|^2+|\Phi_2|^2)$.
However, we will be mostly interested in the zero modes of the hypermultiplets.
 As is well known, at most one fields $\Phi_{1,2}$ can have a zero mode. The $N=1$ supersymmetry acts in such a way that $\Phi_+\equiv\Phi_1$ and $\Phi_-\equiv\bar\Phi_2$ are chiral superfields.~\footnote{Such non-holomorphic relations are typically necessary in order to express quaternionic Kahler metrics in terms of a Kahler potential such as Eq.~(\ref{kahleruni}), see Ref.~\cite{deWit:2001dj}.}
 Truncating to zero modes only, both cases can be described by a Kahler potential:
\be
K_p=-\frac{p}{\kappa}\,\ln\left(1-\frac{\kappa}{p}\,|\Phi_0|^2\right)
\ee
with $p=1$ ($p=2$) for the unitary (symplectic) cosets respectively, and $\Phi_0$ denotes the chiral zero mode that can either belong to $\Phi_-$ or $\Phi_+$. This corresponds to the subspace $\frac{U(1,n)}{U(1)\times U(n)}$. Notice however that the two spaces yield different curvatures $\sim \frac{1}{p}$.

Finally, the gauging generates potentials for the hypermultiplets. In the symplectic case, one has
\begin{multline}
V_S(\Phi)=-12\, k^2\,M^3+F(\Phi)^{-1}\left(m_{\Phi^i_+}^2\,|\Phi^i_+|^2+m_{\Phi_-^i}^2\,|\Phi^i_-|^2\right)\\
-\frac{3\,k^2}{16\,M^3}F(\Phi)^{-2}
\left(|\Phi^j_+|+|\Phi^j_-|^2\right)\left((q^{R}_{\Phi_+^i})^2\,|\Phi_+^i|^2+(q^{R}_{\Phi_-^i})^2\,|\Phi_-^i|^2
\right)
\label{potS}
\end{multline}
The bulk masses are again related to the 5d $R$ charges:
\be
m_{\Phi^i_\pm}^2=\left(c_i^2\pm c_i-\frac{15}{4}\right)k^2\,,\qquad q^R_{\Phi^i_\pm}=1\mp\frac{2}{3}\,c_i
\ . \ee
There are also boundary masses generated \cite{tony}. The boundary condition sets to zero either $\Phi_+$ or $\Phi_-$. The non vanishing field has a boundary mass
\be
\mathcal L^{bd}=-m^{(0)}_{\Phi^i_\pm}\,|\Phi^i_\pm|^2\,\delta(z-z_0)+m^{(1)}_{\Phi^i_\pm}\,|\Phi^i_\pm|\,^2\delta(z-z_1) \ ,
\ee
which are given by
\be
m^{(\alpha)}_{\Phi^i_\pm}=\left(\frac{3}{2}\mp c_i\right)k\,.
\ee

\section{Integrating out 5d gravity in a slice of $AdS$}
\label{sec:kk}

The five-dimensional supergravity Lagrangian in the previous section gives a contribution to the Kahler potential even in the limit of infinite KK masses. There is however another contribution, needed for the consistency of the theory, coming from
integrating out at tree-level massive KK states of the 5d gravitational multiplet. To write down this contribution, it is enough
to write down the linearized bosonic supergravity Lagrangian, coupled linearly to matter.
We therefore show explicitly how to integrate out the heavy KK modes of the graviton
propagating in a slice of $AdS_5$. The results we derive have a more general applicability to non-supersymmetric RS scenarios, generalizing corresponding results from integrating out gauge fields \cite{Hirn:2007bb,Cabrer:2011fb} or fermions \cite{Cabrer:2011qb}.

A massive graviton has two helicity-two, two helicity-one and one helicity-zero degree of freedom which couple to different components of the 5D energy momentum tensor. On top of that there will the trace of the metric, which does not correspond to a physical degree of freedom. We will first show how one can add a Faddeev-Popov gauge fixing term that neatly disentangles the various spins. Let us thus start with the 5d Lagrangian
\be
\mathcal L \ = \ \sqrt{-g} \left[ -M^3(R+\Lambda)-\frac{1}{4}F_{MN}\,F^{MN}+\mathcal L_{\rm matter} \right] \ . \label{start}
\ee
with $\Lambda=-12k^2$ is the negative cosmological constant leading to the AdS$_5$ vacuum.
As usual, we split the metric into a background piece and fluctuations:
\be
g_{MN} \ = \ \gamma_{MN}+\sqrt{\frac{2}{M^3}}\, h_{MN} \ .
\ee
Expanding Eq.~(\ref{start}) to quadratic order in the fluctuations one obtains
(see, e.g., Ref.~\cite{Boos:2002hf})
\begin{multline}
\mathcal L^h/\sqrt{-\gamma} \ = \ - \frac{1}{2}  \biggl(
\nabla_Rh_{MN}\nabla^Rh^{MN}-\nabla_Rh\nabla^Rh
+2\nabla_Mh^{MN}\nabla_Nh\biggr.\\\biggl.
-2\nabla_Mh^{MN}\nabla^Rh_{RN}
\biggr)
+h_{MN}^2+h^2
+\frac{1}{\sqrt{2\,M^3}}h^{MN}T_{MN} \ ,
\end{multline}
where all quantities are covariant with respect to the background metric $\gamma_{MN}$
and the 5d background-covariant energy momentum tensor is defined as
\be
T_{MN} \ = \ -2 \frac{\delta \mathcal L_{\rm matter}}{\delta \gamma_{MN}}+\gamma_{MN}\mathcal L_{\rm matter} \ .
\ee
Notice that the energy momentum tensor in general also contains boundary pieces.
Now decovariantize the action, using the AdS metric in conformally flat coordinates
$
\gamma_{MN} \ = \ (kz)^{-2} \ \eta_{MN}\,.
$
The UV and IR branes are located at $z_0=k^{-1}$ and $z_1$ respectively.
In the following we work in units of the 5d curvature, $k=1$, unless otherwise stated.
Defining $\hat h_{MN}\equiv z^2h_{MN}$ one can write the resulting Lagrangian
\begin{multline}
\mathcal L^h
=
-\frac{1}{2}z^{-3}\left(\partial_R \hat h_{MN}\right)^2
+\frac{1}{4}z^{-3}\left(\partial_N\hat h\right)^2
+3z^{-5}\hat h_{5N}^2
+\frac{z^{-3}}{\sqrt{2\,M^3}}\hat h_{MN}T_{MN}
\\
+z^{-3}\left(
\partial_M \hat h_{MN}-\frac{1}{2}\partial_N\hat h-3z^{-1}\hat h_{N5}
\right)^2 \,.
\end{multline}
Here all contractions are done with $\eta_{MN}$, so no $z$ dependence is left implicit.
The term in the second row is canceled by an appropriate Faddeev-Popov gauge fixing term and can be dropped.
Explicitely writing the different spins one finds for the rest
\begin{multline}
\mathcal L^h \ = \ - \frac{1}{2}z^{-3}(\partial_R \tilde h_{\mu\nu})^2
-\frac{1}{2}z^{-1}(\partial_R B_\mu)^2
+\frac{1}{2}z^{-3}(\partial_R \chi)^2
-\frac{1}{2}z(\partial_R \phi)^2
\\
+\frac{z^{-3}}{\sqrt{M^3}}\left(\frac{1}{\sqrt{2}}\, \tilde h_{\mu\nu}\tilde T_{\mu\nu}
+ z\, B_{\mu}T_{\mu 5}
+ \frac{1}{2\,\sqrt{2}}\chi T_{\mu\mu}
+ \frac{1}{\sqrt{3}}\,z^2\phi\tilde T_{55}\right)
\label{Lag}
\end{multline}
where $\hat h_{\mu5}\equiv z B_\mu/\sqrt{2}$ and we denote by $\tilde h_{\mu\nu}$ the traceless part of $\hat h_{\mu\nu}$. The two scalar degrees of freedom have been disentangled by defining $\chi=\frac{1}{2}(\hat h_{\mu\mu}+2\hat h_{55})$, and $\phi=\sqrt{3/2}\,z^{-2}\hat h_{55}$. The tilded sources are defined as
\be
\tilde T_{\mu\nu}=T_{\mu\nu}-\frac{1}{4}\,\eta_{\mu\nu}\,T_{\rho\rho}\,,\qquad
\tilde T_{55}=T_{55}-\frac{1}{2}\,T_{\rho\rho}\,.
\ee
Focusing on the zero mode for $\hat h_{\mu\nu}$ (which has a constant wave function given by $\sqrt{2}/(z_0^{-2}-z_1^{-2})^{-\frac{1}{2}}$ in our normalization), one deduces the well known relation \cite{rs1} (restoring $k$)
\be
kM_{P}^2 \ = \ M^3 \ (1 - \epsilon^2)\, .
\label{RSPlanck}
\ee
Here $M^2_{P}=(8\pi G_N)^{-1}$ is the reduced Planck mass and $\epsilon = z_0/z_1$.

The Lagrangian in Eq.~(\ref{Lag}) is the starting point for the calculation of the effective action.
%
%
Formally we can write
\bea
\mathcal L^{\rm KK - graviton}_{\rm eff}&=&\frac{1}{4\,M^3}\int_{z_0}^{z_1} dz\,dz' \ z^{-3}\,z'^{-3}\
\tilde T_{\mu\nu}(x,z)\ G_{2}(z,z';-\partial_\mu^2)\ \tilde T_{\mu\nu}(x,z') \nn\\
   &&  -\frac{1}{16\, M^3}\int_{z_0}^{z_1} dz\,dz' \ z^{-3}\,z'^{-3}\
 T_{\mu\mu}(x,z)\ G_{2}(z,z';-\partial_\mu^2)\  T_{\mu\mu}(x,z')
   \nn\\
  &&  +\frac{1}{2\,M^3}\int_{z_0}^{z_1} dz\,dz' \ z^{-2}\,z'^{-2}\
 T_{\mu5}(x,z)\ G_{1}(z,z';-\partial_\mu^2)\  T_{\mu5}(x,z')
   \nn\\
   &&  +\frac{1}{6\, M^3}\int_{z_0}^{z_1} dz\,dz' \ z^{-1}\,z'^{-1}\
 \tilde T_{55}(x,z)\ G_{0}(z,z';-\partial_\mu^2)\ \tilde T_{55}(x,z') \ ,
 \label{Leffgraviton}
\eea
where we have defined the propagators
\be
G_{s}(z,z';p^2)\equiv\sum_n\,\frac{f^n_{s}(z)f^n_{s}(z')}{p^2+m_n^2} \ ,
\ee
which are calculated in App.~\ref{wfprop}, see Eq.~(\ref{solutionprop}). Here the $f_{s}^n$ are the normalized Kaluza Klein wave functions of the various spins of the graviton appearing in Eq.~(\ref{Lag}) where $s=2$ refers to $\tilde h_{\mu\nu}$ and $\chi$, $s=1$ to $B_\mu$, and $s=0$ to $\phi$. In the case $s=0,2$ the propagators contain zero modes which correspond to the massless 4d graviton and the radion. These poles should be subtracted in Eq.~(\ref{Leffgraviton}) in order for $L_{\rm eff}$ to remain local.

The zero mode of $\phi$, the radion, has a constant profile. 
At linear order the radion thus couples to the 4d operator $\int_{z_0}^{z_1} z^{-1}\,\tilde T_{55}(x,z)$.

Moreover, the Lagrangian for the fluctuations of the graviphoton $A_M$ can be obtained from Eq.~(\ref{start}) as:
\be
\mathcal L^{A}=-\frac{1}{2}\,z^{-1}\,(\partial_M A_N)^2+\frac{1}{2}\,z^{-3}\,A_5^2+\frac{1}{2}\,z^{-1}\,\left(\partial_N A_N -z^{-1}A_5\right)^2+z^{-3}A_N\,J_N
\label{LA}
\ee
where
\be
J_N=\frac{\delta \mathcal L_{\rm matter}}{\delta\,A^N}
\ee
The longitudinal part (third term in Eq.~(\ref{LA})) can again be removed by Faddev-Popov gauge fixing. Defining $\rho=z^{-1}A_5$ one gets
\be
\mathcal L^{A}=-\frac{1}{2}\,z^{-1}\,(\partial_M A_\nu)^2
-\frac{1}{2}\,z\,(\partial_M\rho)^2+z^{-3}A_\nu\,J_\nu+z^{-2}\,\rho\,J_5 \ .
\label{LA2}
\ee
Accordingly, one finds
\bea
\mathcal L^{\rm KK - graviphoton}_{\rm eff}&=&  \frac{1}{2}\int_{z_0}^{z_1} dz\,dz' \ z^{-3}\,z'^{-3}\
 J_\mu(x,z)\ G_{1}(z,z';-\partial_\mu^2)\  J_\mu(x,z')
   \nn\\
   &&  +\frac{1}{2}\int_{z_0}^{z_1} dz\,dz' \ z^{-2}\,z'^{-2}\
  J_5(x,z)\ G_{0}(z,z';-\partial_\mu^2)\ J_5(x,z') \ .
 \label{Leffgraviphoton}
\eea

In the next subsection we will apply this to supergravity coupled to hypermultiplets and calculate the contribution to the Kahler potential that can mediate supersymmetry breaking.
\subsection{Contributions to four-fermion operators.}

In this section we give the results for the non-derivative four-fermion terms induced by the exchange of the KK supergravity multiplet.
Adding to it the direct contribution from the 5d Lagrangian allows to calculate the full corrections to the Kahler potential
for matter fields. Most of the details, including the matching with the corresponding bosonic terms, are relegated to the Appendix.
We discuss explicitly the symplectic quaternionic hypermultiplet case and mention at the end of the section the results for the unitary
case. Let us define the following wave functions
\be
f_i(z)=\sqrt{\frac{1-2c_i}{z_1^{1-2c_i}-z_0^{1-2c_i}}}\ z^{\frac{3}{2}-c_i}
\label{scalar}
\ee
which are the normalized wave functions for the scalar zero modes in $\Phi_+$. Their fermionic superpartners $\psi_L$ have wave functions given by $\sqrt{z}\,f_i(z)$. We will only give the result for zero modes contained in $(\Phi_+,\psi_L)$. The result for zero modes sitting in $(\Phi_-,\bar \psi_R)$ can be obtained trivially by making the substitution $c\to-c$. 
%

\subsubsection*{Direct Contribution}

This is the contribution present directly in five dimensions:
\begin{multline}
\mathcal L_{direct}=-\frac{1}{32\,M^3}\int z^{-5} \ (\bar\Psi_i \gamma_\mu\gamma_{5}\Psi_i)^2
=\frac{1}{16 M^3}\int  z^{-3}\,f_i^2f_j^2\ \mathcal O_{ij}\\
=\left[\frac{1}{8M_{P}^2}
+\frac{1}{4}\alpha_{ij}
\right]\mathcal O_{ij} \ , \label{direct1}
\end{multline}
where we defined the operator
\be
O_{ij}(x) \ = \ - \frac{1}{2}(\bar\Psi_i\gamma_\mu\Psi_i)(\bar\Psi_j\gamma_\mu\Psi_j)=
	\psi_i\psi_j\ \bar\psi_i\bar \psi_j \ ,
\ee
where we converted from Dirac to Weyl notation, and the quantity
\be
 \alpha_{ij} \equiv
 \frac{1}{4\,M_{P}^2\epsilon^2}
\frac{(1-2c_i)(1-2c_j)}{(4-2c_i-2c_j)}\frac{(1-\epsilon^{3-2c_i})(1-\epsilon^{3-2c_j})}{(1-\epsilon^{1-2c_i})(1-\epsilon^{1-2c_j})}
-\frac{1}{4\,M_{P}^2}\frac{(3-2c_i)(3-2c_j)}{(4-2c_i-2c_j)} \ .  \label{alpha2}
\ee
\subsubsection*{Contribution from $A_\mu$.}
The linear coupling between the matter fields and the graviphoton $A_{\mu}$ in Eq.~(\ref{LA2}) can be derived from Eq.~(\ref{Lfermion}) and is given by
\be
\int z^{-3} A_\mu\,J_\mu=
g_R\sum_i\int z^{-4}\bar\Psi_i(x,z)\left[\left(-\frac{2}{3}\,c_i\right)\gamma_\mu\, A_\mu(x,z)+\frac{1}{2}\,z\,\gamma_\mu\gamma_5\,\partial_z A_\mu(x,z)\right]\Psi_i(x,z)\,.
\ee
This has to be used in Eq.~(\ref{Leffgraviphoton}). Using that the zero modes are $\Psi(x,z)=\sqrt{z}f_i(z)\Psi(x)$ with $f_i$ given in Eq.~(\ref{scalar}) one gets
\be
 J_\mu(x,z)=
 -\frac{g_R}{2}\sum_i\left[1- \frac{2}{3}c_i\right]f_i^2(z)
	\ \bar\Psi_i(x)\gamma_\mu\Psi_i(x)\,.
\ee
%
Since we are only interested in the zero-derivative terms, we can proceed the evaluation as described in App.~\ref{zeromomentum}.
For this we need to compute
\bea
\mathcal J_\mu(x,z)&=&\int_{z_0}^z z'^{-3}J_\mu(x,z')
\nn\\
&=&-\frac{g_R}{2}\sum_i\left[1- \frac{2}{3}c_i\right]\Omega_{1- 2c_i}(z)
	\ \bar\Psi_i(x)\gamma_\mu\Psi_i(x)
	\label{Jfermion}
\eea
with the functions
\be
\Omega_\alpha \ = \ \frac{z^\alpha-z_0^\alpha}{z_1^\alpha-z_0^\alpha}\,.
\label{Omegadef}
\ee
%
%
The contribution of the KK exchange of the graviphoton is then, using Eq.~(\ref{Jfermion}) in Eq.~(\ref{Leffgraviphoton2})
\bea
&& {\cal L}^A 
= \frac{3}{32\,M^3}
\left(1-\frac{2}{3}c_i\right)\left(1-\frac{2}{3}c_j\right)
\left[
\int_{z_0}^{z_1} z\ \Omega_{1- 2c_i}\Omega_{1- 2c_j}\right.
\nn\\
&&\qquad \qquad \left.-\frac{2}{z_1^2-z_0^2}\int_{z_0}^{z_1} z\ \Omega_{1- 2c_i}\int_{z_0}^{z_1} z\ \Omega_{1- 2c_j}
\right] (-2\mathcal O_{ij}) \ .
\eea
The integrals can be evaluated straightforwardly and the result is seen to be proportional to $\alpha_{ij}$ defined in Eq.~(\ref{alpha2}):
\be
{\cal L}^A
= \ - \ \frac{1}{12}\alpha_{ij}\mathcal O_{ij} \ .
\ee
%
\subsubsection*{Contribution from $\tilde h_{\mu\nu}$, $B_\mu$, $\chi$ and $\phi$.}

The 5d energy momentum tensor for fermions reads
\begin{multline}
T_{MN}=\frac{i}{4}z^{-1}\left(
\bar \Psi \gamma_M \Psi_{;N}+\bar\Psi\gamma_N \Psi_{;M}-\bar \Psi_{;N}\gamma_M\Psi-\bar\Psi_{;M}\gamma_N\Psi
\right)\\
-\frac{i}{2}z^{-1}\eta_{MN}(\bar\Psi\gamma_S\Psi_{;S}-\bar\Psi_{;S}\gamma_S\Psi)
-i\eta_{MN}z^{-2}m_\Psi\bar\Psi\Psi \ ,
\end{multline}
where the semicolon denotes covariant differentiation $\partial_M+\Gamma_M$ with $AdS_5$ spin connection \cite{tony}
\be
\Gamma_\mu=\frac{1}{2}z^{-1}\gamma_5\gamma_\mu\, \quad , \quad \Gamma_5 = 0 \, .
\ee
Since the zero modes of the field $\Psi$ are chiral, it is clear that we can drop all terms that
contain one left and one right handed spinor. One can then immediately verify that all remaining terms in  $T_{\mu\nu}$ and $T_{55}$ contain $\partial_\mu$ derivatives and hence do not contribute.
It is straightforward to check that the nonderivative contributions to $T_{\mu 5}$ vanish for the fermionic zero modes by making use of the explicit form of the spin connection.
 Therefore
there is no contribution from the $B_{\mu}$ exchange either on the four-fermion terms.

A comment is in order regarding the possible contribution from a stabilized (massive) radion.
The fact that $\phi$ only couples derivatively to chiral fermion zero modes applies in particular to the radion, $\phi^0(x)$. We thus do not expect any tree-level contribution to the Kahler potential coming from the radion once it acquires a mass.

\subsection{The effective quartic Kahler potential}
\label{sec:quartic}

The sum of all contributions gives the four-fermion term
\be
\mathcal L_{\rm eff}^{4f}=-\frac{1}{8\,M_{P}^2}\mathcal \sum_{ij}\mathcal O_{ij}+\frac{1}{4\,M_{P}^2}\mathcal \sum_{ij}\mathcal O_{ij}+\frac{1}{6}\sum_{ij}\alpha_{ij}\,\mathcal O_{ij} \ ,
\label{fullfermion}
\ee
where in the $ij$ independent term we have extracted a term that should be attributed to the metric rather than to the curvature (i.e.~the quartic Kahler terms):~\footnote{We use the common convention $R_{i \bar j k \bar l} =K_{i \bar j k \bar l} -g^{m\bar n}K_{\bar j m\bar l}K_{i\bar n k}$ which yields $R>0$ for negatively curved spaces such as $\frac{U(1,n)}{U(1) \times U(n)}$.}
\be
\mathcal L^{4f}_{\rm eff}=\sum_{ij}\left(-\frac{1}{8\,M_P^2}g_{i\bar j}g_{k\bar l}+\frac{1}{4}R_{i\bar j k\bar l}\right)\psi_i\,\psi_k\,\bar \psi_{\bar j}\,\bar\psi_{\bar l}
\ee

One sees that  Eq.~(\ref{fullfermion}) corresponds to the following quartic term in the matter fields in the 4d effective Kahler potential
\be
K^{(4)} = \sum_{ij }\left( \frac{1}{4\,M_{P}^2} + \frac{1}{6}\, \alpha_{ij}  \right) |\Phi_i|^2 |\Phi_j|^2 \ , \qquad {\rm symplectic \ case} \ . \label{full1}
\ee
The results for the unitary case are very similar. The only difference is in the direct contribution in 5d, which is three times
bigger than in (\ref{direct1}), as transparent in (\ref{4fS}), (\ref{4fU}). This results in an effective quartic Kahler potential
\be
K^{(4)} = \sum_{ij }\left( \frac{1}{2\,M_{P}^2} + \frac{2}{3}\, \alpha_{ij}  \right) |\Phi_i|^2 |\Phi_j|^2 \ , \qquad {\rm unitary \ case} \ . \label{full2}
\ee
Before turning to compute the scalar masses, let us comment on a few features of the result. Besides the  parameters $c_i$, the $\alpha_{ij}$ only depend on the UV and IR scales $M_P$ and $\epsilon\,M_P$. Which one of the two terms in Eq.~(\ref{alpha2}) dominates depends on the relative values of $c_i$ and $c_j$.
We will show below, see Eq.~(\ref{alphamore}), that there is a lower bound on $\alpha_{ij}$ for any pair $c_i,\ c_j$ which thus cannot become arbitrarily negative.
An interesting feature of the result is that it diverges for both $\Phi_i$ and $\Phi_j$ strongly localized towards the same boundary, e.g.~$c_i,c_j\to \infty$. The reason is that the 5d couplings such as the 5d $R$ charge, Eq.~(\ref{masscharge}), diverge in this limit.

The quartic Kahler extracted here from the zero mode Lagrangian of the fermions can equally well be calculated by looking at quartic two-derivative operators of their scalar superpartners. As it turns out, besides direct contributions and contributions from $A_\mu$ there are now nonzero contributions from $B_\mu$, $\chi$ and $\phi$ to these operators. The computation is carried out in App.~\ref{app:scalars}. The generated scalar operators result in the same Kahler potential, Eqns.~(\ref{full1}) and (\ref{full2}) as the four fermion terms. This is a nontrivial consistency check of our result.
An important point we would like to stress here is that the direct contributions (5d four-fermion terms vs.~5d sigma model terms) separately do not give a supersymmetric result. Only after the exchange of KK modes of the supergravity multiplet is taken into account do the results agree. A naive truncation to zero modes is thus not correct. The only case where it works is in the infinite KK mass limit. This occurs when $z_1\to z_0$ or equivalently $\epsilon\to 1$. The KK exchange is always proportional to $\alpha_{ij}$ and in this limit $\alpha_{ij}\to 0$. The $\alpha_{ij}$ independent terms in the direct contributions then do combine into a supersymmetric result.

Another interesting cross-check is the computation of the 4d scalar potential. The scalar zero mode sector has a $G_0=U(1)_R'\times U(n_{c_1})\times U(n_{c_2})\times\dots\subset U(1)_R'\times U(n)$ global symmetry~\footnote{The subgroup $U(1)_R'$ is the subgroup of the $SU(2)_R$ global symmetry.
The $n_{c_i}$ denote the number of chiral superfields with localization parameter $c_i$.
 The subgroup $U(1)_R$ gauged by the graviphoton is the diagonal of all abelian factors in $G_0$.}  which rules out an effective superpotential and hence an effective scalar potential.
The latter has a priori four contributions: The direct terms from Eq.~(\ref{potS}), the direct contributions from Eq.~(\ref{kahlersym}) with extra-dimensional derivatives, as well as contributions from exchange of KK modes of $\phi$ and $\chi$.
The mass terms cancel between the two direct contributions (a simple consequence of the fact that the zero modes are exactly massless). Furthermore we have checked that the quartic potential terms all cancel in a nontrivial way between the four contributions mentioned above. We expect this cancelation to hold for higher order terms but we have not checked this explicitly.


\section{Scalar masses}
\label{sec:scalar}

The starting point in the computation of visible sector scalar masses is the Kahler potential obtained after the reduction
to 4d \footnote{In what follows, we the the convention $M_P=1$.}
\be
K \ = \ - p \ \ln \left(1 - \frac{1}{p} \sum_{i=1}^n |\Phi_i|^2\right) \ + \ d_p \alpha_{ij} |\Phi_i|^2 |\Phi_j|^2 \ , \label{sc1}
\ee
where $p = 1$ , $d_{p=1} \ = \ \frac{2}{3}$  { for the  unitary quaternionic spaces}
$\frac{U(2,n)}{U(2) \times U(n)}$ and
$p = 2$ , $d_{p=2} \ = \ \frac{1}{6}$  { for the symplectic quaternionic spaces}
$\frac{USp(2,2n)}{USp(2) \times USp(2n)}$.
Notice that that the first term in (\ref{sc1}), parameterizing the coset space $\frac{U(1,n)}{U(1) \times U(n)}$, describes the truncation of both quaternionic spaces to ${\cal N}=1$ in 4d. The quartic terms in its expansion reproduce the first (alpha-independent) terms
in the r.h.s.~of (\ref{full1}) (for $p=2$) and (\ref{full2}) (for $p=1$). It correspond to the $\alpha =0$ case and can be understood as the
$\epsilon \rightarrow 1$ ($M_P$ held fixed) limit in which the KK masses decouple. This part can be obtained exactly from the 5d theory, beyond the quartic term
displayed in the previous section.\footnote{In the denominators of the sigma model metric the limit gives $M^{-3} f_i^2\to 2\,M_{P}^{-2}$ using Eq.~(\ref{RSPlanck}).}
The $\alpha$ dependent term is the lowest term coming from the integration of heavy states. This is in principle only the first term in
an expansion in number of matter fields; higher-order terms are expected to be induced.

Scalar soft masses (for unnormalized kinetic terms) for visible matter fields are computed starting from \cite{soft}
\be
m_{a \bar b}^2 \ = \ m_{3/2}^2 \ (G_{a \bar b} - G^{\alpha} R_{a {\bar b} \alpha {\bar \beta}} G^{\bar \beta}) \ , \label{sc3}
\ee
where indices $a,b$ stand for visible matter fields and $\alpha,\beta$ for SUSY breaking fields with
$G^{\alpha} G_{\alpha \bar \beta }G^{\bar \beta} = 3$. By using the fact that the Kahler potential in the first term
in (\ref{sc1}) describes an Einstein space with\be
R_{i \bar j k \bar l} \ = \ \frac{1}{p} (G_{i \bar j} G_{k \bar l} + G_{i \bar l} G_{k \bar j} ) \ , \label{sc4}
\ee
it is then easy to check that this geometric part contributes, after normalization of the kinetic terms
\be
(m_a^2)_{\alpha_{ij}=0} \ = \ m_{3/2}^2  \left(1 - \frac{3}{p}\right) \ .  \label{sc5}
\ee
The second term, dependent on the localization of fields, equals
\bea
&& (m_a^2)_{\alpha_{ij}}  =  - \frac{4\alpha_{a \beta} }{3} |F_{\beta}|^2  =  -
\frac{4m_{3/2}^2}{3}  \alpha_{a \beta} |G_{\beta}|^2 \ , \qquad {\rm unitary \ case \  }p=1 \ , \nonumber \\
&& (m_a^2)_{\alpha_{ij}} =  - \frac{\alpha_{a \beta}}{3}  |F_{\beta}|^2  =  -
\frac{m_{3/2}^2}{3}  \alpha_{a \beta} |G_{\beta}|^2 \ , \qquad {\rm symplectic \ case \  }p=2 \ ,
\label{alphadependent}
\eea
where $|G_{\alpha}|^2 = G_{\alpha \bar \beta} G^{\alpha} G^{\bar \beta}$.
Putting the two terms together, we find the scalar soft masses
\bea
&& m_a^2 \ = \ - 2 m_{3/2}^2 \left(1 + \frac{2}{3} \alpha_{a \beta} |G_{\beta}|^2\right) \qquad {\rm  unitary \ case \ }p=1 \ , \nonumber  \\
&& m_a^2 \ = \ - \frac{1}{2} m_{3/2}^2 \left (1 + \frac{2}{3} \alpha_{a \beta} |G_{\beta}|^2\right) \qquad
{\rm symplectic \ case \ }p=2 \ .
\label{sc6}
\eea
Notice that
\be
1 + 2 \alpha_{ij}=
\frac{(1-2c_i)(1-2c_j)}{2 (4-2c_i-2c_j)} \left[ \frac{(1-\epsilon^{3-2c_i})(1-\epsilon^{3-2c_j})}{\epsilon^2 (1-\epsilon^{1-2c_i})(1-\epsilon^{1-2c_j})}
- 1 \right] \ \geq \ 0 \ , \label{alphamore}
\ee
the equality corresponding to the "sequestered" case $c_i = - c_{\beta} = \pm \infty$ where the matter field $\Phi_a$ and the SUSY
spurion $\Phi_{\beta}$ sit at the opposite boundaries of the internal space $S^1/Z_2$. Precisely in this case, by using the cancelation
of the cosmological constant  $|G_{\alpha}|^2 = 3$ the visible sector scalar masses vanish! The cancelation in  this sequestered case
was actually argued in more general terms in \cite{rs3}. This is therefore a non-trivial consistency check of our computation and framework. In particular, the effective operators induced by the KK states were crucial in order to get the agreement with sequestering. On the other hand $\alpha$'s are bounded from below in (\ref{alphamore}); in all other situations we find therefore that the scalar masses that we computed are {\it tachyonic}.  Since the contribution to scalar masses that we computed is model-independent, coming from the irreducible gravitational multiplet, this put strong constraints on model building. It implies that new contributions to the effective Kahler potential have to be present in order to avoid vacuum instability:
\begin{itemize}
\item One obvious possibility are brane localized Kahler potentials. Below we estimate the coefficients and show that they have to be quite large, making it an unlikely solution to the problem.
\item Another possibility is that the hypermultiplet containing the spurion is charged under some other 5d gauge symmetry. The KK modes of such extra vector multiplets do not come with the dangerous contact interactions that are responsible for the tachyonic masses, and, hence, can lead to positive contributions to the soft masses squared. Interestingly, the associated 5d gauge coupling $g_{5d}$ can be somewhat larger than the graviphoton gauge coupling, $g_{5d}^2>g^2_R=3k^2/4M^3$ while still maintaing 5d perturbativity.
The reason is that while $g_{5d}$ is just bounded by demanding loop corrections to be small, $g_R$ is also bounded by demanding higher curvature corrections to be suppressed, $k\ll M$. We will discuss this option in detail in the next section.
\item
Another possibility is gauge mediation at a scale lower than $\epsilon\,M_P$.
\item
SUSY breaking
could also have contributions from other $F$ terms, such as the  radion, bulk vector multiplets or exactly brane-localized fields. Our bulk spurion(s) do not need to saturate the cosmological constant in this case, $|G_\alpha|^2<3$.  However, the $\alpha$ dependent contributions Eq.~(\ref{alphadependent}) are still tachyonic for $\alpha_{\alpha j}>0$, while for $\alpha_{\alpha j}<0$ their positive contribution decreases.
\end{itemize}

Let us briefly estimate the coefficients that brane operators need to have in order to cancel the bulk contribution.
A quartic Kahler potential of bulk fields localized on the brane yields dimension-8 operators, and hence
it should be suppressed by four powers of the 5d scale $M$. Inserting the wave functions and the induced metric one finds that the quartic Kahler operators on the brane have coefficients
\bea
\beta^{UV}_{ij}&=&b^{UV}\,\frac{k^2}{M^4}\frac{(1-2c_i)(1-2c_j)}{(1-\epsilon^{1-2c_i})(1-\epsilon^{1-2c_j})}\epsilon^{1-2c_i}\epsilon^{1-2c_j}\nn\\
\beta^{IR}_{ij}&=&b^{IR}\,\frac{k^2}{M^4\,\epsilon^2}\frac{(1-2c_i)(1-2c_j)}{(1-\epsilon^{1-2c_i})(1-\epsilon^{1-2c_j})}
\eea
where we have assumed the dimensionless coefficients $b^{UV,\,IR}$ to be universal for simplicity.
Comparing these expressions with the tachyonic contributions one can see that the scaling with $\epsilon$ is precisely the same in the two cases.~\footnote{In fact the $\epsilon$-scaling in the $c_i$-$c_j$ plane is the same as that found in Sec.~\ref{sec:vector} coming from integrating out an external vector multiplet with $(+,-)$ boundary conditions, see Fig.~\ref{figure2}.}
However, using $M^3 \simeq M_P^2\,k$, we see that the brane localized terms carry an additional factor of $k/M$.  This quantity should be a somewhat small number in order to ensure perturbativity of the supergravity theory, implying that the dimensionless coefficients have to be quite large in order to overcome the irreducible  tachyonic contribution from the bulk.
Engineering models of supersymmetry breaking with a viable mass spectrum can thus become a difficult task in this minimal setup.
\section{Additional bulk vector fields and phenomenology}
\label{sec:vector}

As has been pointed out in the previous section, we expect to be able to resolve the issue of tachyonic soft masses by integrating out Kaluza Klein modes of additional gauge symmetries present in the 5d bulk, provided the susy breaking spurion is charged.
If these gauge symmetries are completely broken by the boundary conditions the scalar superpartners and fifth component of the gauge bosons form part of the 4d sigma model after compactification. This case has recently been studied in Ref.~\cite{abe}. Here we will focus on the case of boundary conditions that project out the moduli, so we do not have to worry about their stabilization. In this section we will follow the orbifold terminology and label the boundary conditions (BC) of the gauge field as $+$ (Neumann) or $-$ (Dirichlet).
We can exclude the presence of zero modes for $A_5$ by choosing the BC for $A_\mu$ to be $(+,+)$, $(+,-)$ or $(-,+)$.

To integrate out the KK modes of these gauge fields we follow the same procedure as for the graviphoton field, replacing the propagator of the latter (which has $(-,-)$ BC) with the one corresponding to the new BC. These propagators have been computed in Ref.~\cite{Cabrer:2011fb} for the case of zero KK momentum.
Up to terms with 4d derivatives on the currents we can write
\bea
\mathcal L^{ A_\mu^{I++}}_{\rm eff}&=&
  \frac{1}{2}\int_{z_0}^{z_1} dz\ z\,\bigl[\,\mathcal J^I_{\mu }(z)-\Omega_0(z)\,\mathcal J^I_\mu(z_1)\,\bigr]^2
           						+\dots \ , \nn\\
\mathcal L^{ A_\mu^{I+-}}_{\rm eff}&=&
  \frac{1}{2}\int_{z_0}^{z_1} dz\ z\,\bigl[\,\mathcal J^I_{\mu }(z)\, \bigr]^2
           						+ \dots \ , \nn\\
\mathcal L^{ A_\mu^{I-+}}_{\rm eff}&=&
\frac{1}{2}\int_{z_0}^{z_1} dz\ z\,\bigl[\,\mathcal J^I_{\mu }(z)-\mathcal J^I_\mu(z_1)\, \bigr]^2
           						+\dots \ \ . 			
\eea
Here $\mathcal J^I$ are defined as in Eq.~(\ref{defcur}) with the 5d graviphoton current $J$ replaced by the corresponding 5d currents $J^I$ coupling to the gauge field $A^I$ (by definition, the $J^I$ contain a power of the 5d gauge coupling). In the first case ($+,+$) the contribution from the zero mode has been subtracted. However, for phenomenological reasons this zero mode will eventually have to become massive. If its mass is below the IR scale it will dominate the quartic interactions. We will thus focus on the two cases $(+,-)$ and $(-,+)$. We remind the reader that the $(-,+)$ case describes an exact global symmetry of the conformal sector, while the $(+,-)$ one a global symmetry of the conformal sector which is gauged by an elementary gauge field but spontaneously broken by the strong dynamics. In both cases the KK masses are set by the curvature $k$, but the couplings to elementary/composite states are quite different.

\begin{figure}[thb]
\begin{center}
\psfrag{L1}[][lc]{$\frac{1}{2}$}
\psfrag{L2}[][cb]{$\frac{1}{2}$}
\psfrag{UVUV}[][cc][1.2]{ $\frac{1}{2\epsilon^2}$}
\psfrag{IRUV}[][cc][1.2]{ $\frac{1}{\epsilon^2}\frac{1}{3-2c_X}$}
\psfrag{IRIR}[][cc][1.2]{$\frac{1}{\epsilon^2}\frac{1}{4-2c_X-2c_Q}$}
\psfrag{UVIR}[][cc][1.2]{ $\frac{1}{\epsilon^2}\frac{1}{3-2c_Q}$}
\psfrag{cX}{$c_X$}
\psfrag{cQ}{$c_Q$}
\epsfig{file=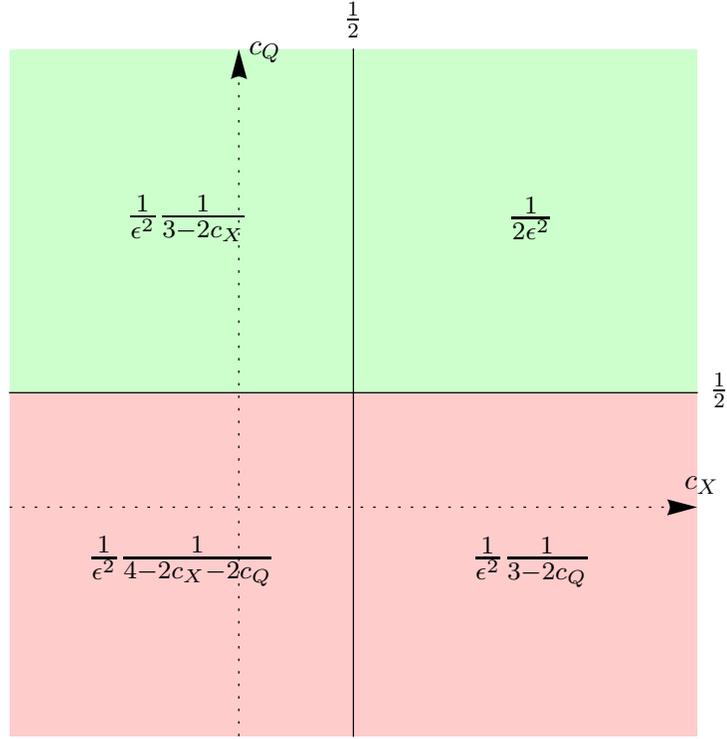,width=10cm}
\end{center}
\caption{Leading contributions to scalar masses as a function of the localization parameters for matter and spurion field, in the case of $(+,-)$ boundary conditions. The upper (green) region for $c_Q>\frac{1}{2}$ yields flavor-universal soft masses.}
\label{fig1}
\end{figure}

Focussing on abelian gauge multiplets, and following our strategy to extract the quartic Kahler from the four-fermion interactions we compute
\be
\mathcal L_{\rm eff}^{A_\mu^{I}} \ = \ - \sum_{i,j}\alpha^I_{ij}\ \mathcal O_{ij} \ ,
\ee
where
\bea
\alpha_{ij}^{I+-}&=&g_I^2\, q^I_i\,q^I_j\ \int_{z_0}^{z_1} dz\,z\ \Omega_{1-2c_i}\,\Omega_{1-2c_j}\nn\\
   &=&  \frac{g_I^2}{k}\, \frac{ q^I_i\,q^I_j}{1-\epsilon^2}\left[
	\frac{1}{4-2c_i-2c_j} \left(\frac{1}{\epsilon^2}R_i\,R_j -1\right)
	+\frac{1}{2\,\epsilon^2}\ (R_i-1)(R_j-1)
   \right]
   \label{plusminus}
\eea
and
\bea
\alpha_{ij}^{I-+}&=&g_I^2\, q^I_i\,q^I_j\ \int_{z_0}^{z_1} dz\,z\ (\Omega_{1-2c_i}-1)\,(\Omega_{1-2c_j}-1)\nn\\
   &=&  \frac{g_I^2}{k}\,\frac{ q^I_i\,q^I_j}{1-\epsilon^2}\left[
	\frac{1}{4-2c_i-2c_j} \left(\frac{1}{\epsilon^2}R_i\,R_j -1\right)
	+\frac{1}{2\,\epsilon^2}\ (R_i-\epsilon^2)(R_j-\epsilon^2)
   \right] \ ,
\eea
where the $q_i^I$ are the charges, $g^I$ the 5d gauge couplings, and
\be
R_i = \frac{(1-2c_i)}{(3-2c_i)}\,\frac{1-\epsilon^{3-2c_i}}{1-\epsilon^{1-2c_i}}
\approx \frac{|1-2c_i|}{|3-2c_i|} \left\{\begin{array}{cc}
1& \ , \ \ c_i<\frac{1}{2}\\
\epsilon^{2c_i-1}& \ \ \ \ \ \ , \ \ \frac{1}{2}<c_i<\frac{3}{2}\\
\epsilon^2& \ , \ \ \frac{3}{2}<c_i
\end{array}\right.
\ee
is an everywhere positive and monotonically decreasing function of $c_i$.
As before we can extract the quartic Kahler potential as
\be
K^{(4)} \ = \ - \sum_{ij } \alpha^{I}_{ij}\  |\Phi_i|^2 |\Phi_j|^2 \ .
\ee
Notice that in the particular case where the spurion and the matter field are both (strongly) localized on the UV brane,
$\alpha_{ij}^{I+-} \rightarrow g_I^2 q^I_i\,q^I_j / 2 \epsilon^2$, whereas $\alpha_{ij}^{I-+}
\rightarrow - g_I^2 q^I_i\,q^I_j / (4-2c_i-c_j)$. We display in Fig.~\ref{fig1} the approximate values of the scalar masses as a function
of the localization parameters $c_Q,c_X$ of the matter and the spurion field, in the case of $(+,-)$ boundary conditions (the expressions correspond to the square bracket in Eq.~(\ref{plusminus})). Of particular interest is the green region in which
scalar masses are universal (for equal $U(1)$ charges), i.e. independent of the localization of the matter fields, provided they are
mostly UV localized. On the other hand, for IR localization of matter fields, scalar masses scale inversely proportional to the degree
of IR localization. The stronger the IR localization, the smaller the corresponding scalar mass\footnote{
The smallness of soft masses for IR localized fields in this configuration is due to the IR Dirichlet BC for the gauge field. Interestingly though, this sequestering is rather weak: in order to suppress the $U(1)$ mediated contributions to the level of gravity mediated contributions, one would need $c_Q\gtrsim \left(M_{P}/k\epsilon\right)^2$.}.
 In phenomenological RS models with an IR localized Higgs field, in order
to explain fermion masses by the various localization of SM fermions, the first two generations are usually UV localized, whereas the third one is IR localized. The $U(1)$ being broken on the IR brane, fermion mass matrices are completely determined by the localization pattern there
\be
Y_{ij}^U \sim
\begin{pmatrix}  \epsilon^{q'_1+u'_1} & \epsilon^{q'_1+u'_2} & \epsilon^{q'_1}  \\ \epsilon^{q'_2+u'_1} & \epsilon^{q'_2+u'_2} &
\epsilon^{q'_2} \\ \epsilon^{u'_1} & \epsilon^{u'_2} & 1 \end{pmatrix}
\quad , \quad Y_{ij}^D \sim
\begin{pmatrix}  \epsilon^{q'_1+d'_1} & \epsilon^{q'_1+d'_2} & \epsilon^{q'_1}  \\ \epsilon^{q'_2+d'_1} & \epsilon^{q'_2+d'_2} &
\epsilon^{q'_2} \\ \epsilon^{d'_1} &  \epsilon^{d'_2} & 1 \end{pmatrix} \ ,
\ee
where $Y_{ij}^U$ ($Y_{ij}^D$) are up-type (down-type) Yukawa matrices, $q'_1 = c_{q_1}-1/2$, etc are analogs of Froggatt-Nielsen charges for the first generation of left-handed quarks $Q_1$ in abelian flavor models \cite{fn}, etc. Unlike usual flavor SUSY models however, in our case there is no conflict between having realistic fermion
 masses and suppression of FCNC effects below present experimental bounds. Combined with the previous comments, this seems to select
an inverted SUSY spectrum, in which the first two generations are heavier than the third one. More precisely, in our case the third generation is lighter than the first two generations which are degenerate. However, a genuine mass hierarchy between the first two
generation scalars versus the third generation scalars, like in the so-called natural SUSY, is obtained only for very strong third generation localization parameters which is beyond the validity of the effective theory approach. Alternatively, universality of scalar masses (like in the constrained version of MSSM, CMSSM) would imply UV localization for all matter fermions. This
is certainly possible, but at the prize of abandoning explaining fermion mass hierarchies through different localization in
the extra dimension for the three generations.

\begin{figure}[thb]
\begin{center}
\psfrag{L1}[][lc]{$\frac{1}{2}$}
\psfrag{L2}[][cb]{$\frac{1}{2}$}
\psfrag{L3}[][cb]{$\frac{3}{2}$}
\psfrag{L4}[][lc]{$\frac{3}{2}$}
\psfrag{UVUV}[][cc][1.]{ $-\frac{1}{4-2c_Q-2c_X}$}
\psfrag{IRUV}[cc][cc][1.]{ $-\frac{1}{3-2c_Q}$}
\psfrag{IRIR}[][cc][1.]{$\frac{1}{\epsilon^2}\frac{1}{4-2c_X-2c_Q}$}
\psfrag{UVIR}[][cc][1.]{ $-\frac{1}{3-2c_X}$}
\psfrag{IR4}[][cb][1.][0]{$\ \ x_Q\, \epsilon^{2c_Q+2c_X-4}$}
\psfrag{IR1}[][cc][1.]{$x_Q\,\epsilon^{-2}$}
\psfrag{IR2}[][cc][1.]{$-x_Q\,\epsilon^{2c_X-3}$}
\psfrag{IR3}[][cc][1.]{$-x_Q\,\epsilon^{2c_Q-3}$}
\psfrag{cX}{$c_X$}
\psfrag{cQ}{$c_Q$}
\epsfig{file=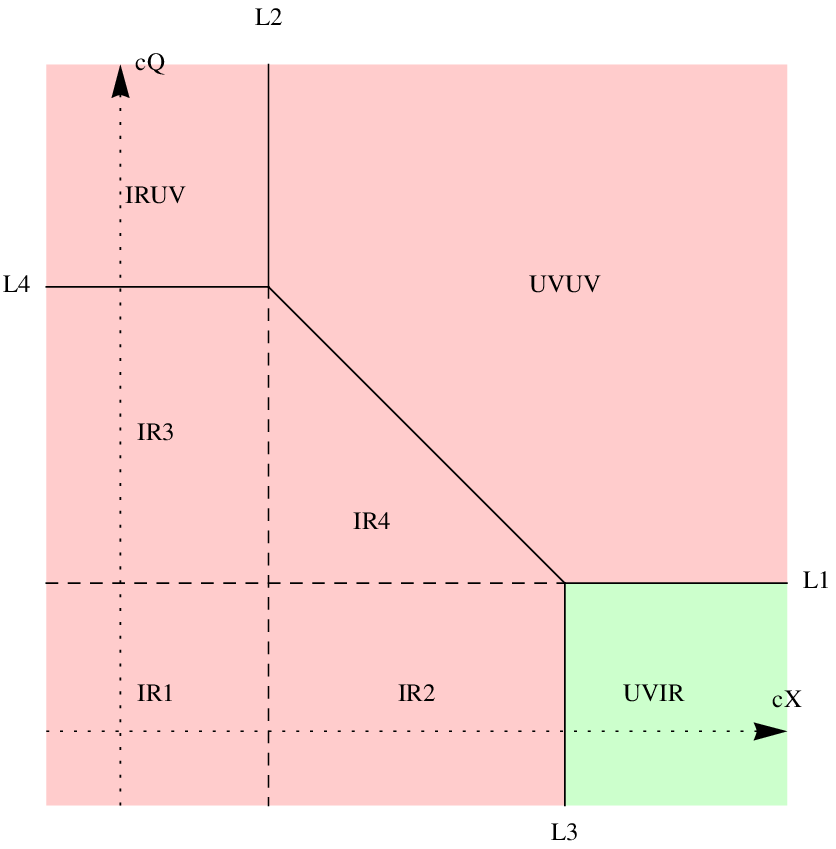,width=10cm}
\end{center}
\caption{Leading contributions to scalar masses as a function of the localization parameters for matter and spurion field, in the case of $(-,+)$
boundary conditions. Here $x_Q=(1-2c_Q)(1-2c_X)(3-c_Q-c_X)/(3-2c_Q)(3-2c_X)(4-2c_Q-2c_X)$. Only the region of $c_X>\frac{3}{2}$, $c_Q<\frac{1}{2}$ is flavor blind.}
\label{figure2}
\end{figure}

There are several comments we can make, by comparing the operators induced by the KK gravity multiplet with the ones generated by the abelian vector multiplets :
\begin{itemize}
\item
The contributions from the $(+,-)$ gauge field scale as $\epsilon^{-2}$ for any values of $c_Q,c_X$.
Contributions to soft terms from 5d supergravity as well as those coming from brane localized Kahler potentials scale as $\epsilon^{-2}$ only for the case if both $c_Q,c_X<1/2$ while in other regions they are more suppressed.
\item Natural values of 5d gauge couplings are $g^2 \sim 1/M$, which are larger than the graviphoton coupling $g_R^2 = 3k^2/4M^3$.
This implies that the dimension six operators generated by the bulk vector multiplet exchange are naturally enhanced by a factor
$(M/k)^2$ compared to the gravity tachyonic contributions. Formally this is true only for not very large values of the localization parameters. However, as already emphasized in Sec.~\ref{sec:quartic}, for strongly localized spurion and matter fields towards the same
brane, the gravity multiplet couplings to matter fields grow and the effective field theory  description breaks down.
\item The strong coupling problem for the couplings to KK gravity multiplet does not arise for the bulk vector multiplets. In particular induced scalar masses are well-defined for any values of the localization parameters.
\end{itemize}

Neither the graviphoton exchange nor the integration of the vector multiplets produces gaugino masses, $A$-terms or the $B \mu$ parameter. These soft terms must thus be present as local operators on the boundaries. In what follows we focus on the $(+,-)$ BC case.
The gaugino masses can originate from linear terms (in $X$) in the gauge kinetic function on the IR brane (the corresponding UV term
 is forbidden by the $U(1)$ symmetry, which is unbroken on the UV brane), which for a bulk spurion is suppressed as $M^{-5/2}$. For instance, for $c_X<1/2$ the main contribution comes from the IR brane
\bea
M_{1/2}=f_{IR}\ \frac{k^\frac{3}{2}\, F_X}{M^\frac{5}{2}\,\epsilon}\ \sqrt{\frac{1-2\,c_x}{1-\epsilon^{1-2\,c_X}}}\
 \frac{1}{\ln\epsilon^{-1}}\,,
\simeq f_{IR}\ \frac{k^\frac{3}{2}\, F_X}{M^\frac{5}{2}\,\epsilon}\ \frac{\sqrt{1-2\,c_x}}{\ln\epsilon^{-1}} \ ,
\eea
where $f_{IR}$ is an $\mathcal O(1)$ number. This means that gaugino masses are suppressed  compared to the soft scalar masses generated from integrating out the vectors as
\be
\frac{M_{1/2}}{m_0} \sim  \frac{k^2}{M^2}\,.
\label{gauginosuppression}
\ee
 Similarly, $A$-terms can result from quartic superpotential terms on the boundaries which for four bulk fields scale as $M^{-3}$. This leads to a relative suppression compared to the scalar masses $\sim (k/M)^{5/2}$. While small $A$ terms can be welcome in view of flavor and CP violation, it also makes it difficult to obtain large mixing in the stop sector needed in order to push the Higgs mass up towards $\sim 125$ GeV. However, this is not a problem here since the stop mass is typically very heavy, as we will see in a moment.

Let us finally comment on the Higgs sector. For IR localized Higgs doublets ($c_{H^{1,2}}<1/2$) one can write explicit operators in the IR brane Kahler potential
\be
K_{\rm IR} \ \supset \ \frac{X^{\dagger } X}{M^4} H_1\, H_2 \ + \  \frac{X^{\dagger }}{M^{5/2}} H_1\, H_2 \ ,
\ee
where the powers of $M$ are determined from the 5d dimensions of the fields, which yields $\mu$ and $B \mu$ terms via the Giudice-Masiero mechanism. One easily figures out the scaling $B \mu/m_{1,2}^2\sim (k/M)^3$ and $\mu^2 /m_{1,2}^2\sim (k/M)^4$ , and hence large $\tan\beta$ is expected. Requiring $\tan\beta\lsim 50$ gives the estimate $k/M\sim 1/4$. This in turn implies that according to Eq.~(\ref{gauginosuppression}) gaugino masses are typically a factor of $\sim 16 $ smaller than the scalar masses.
We considered until now the case of IR localized spurion ($c_X < 1/2$). This turns actually to be the only realistic case, since
for UV spurion localization gaugino masses and Higgs mass parameters have an additional suppression, enhancing further the hierarchy
between the gaugino and scalar masses and increasing further the value of $\tan \beta$.

The case of $(-,+)$ boundary conditions yields a qualitatively different picture. The scaling and flavor structure of the resulting soft terms are displayed in Fig.~\ref{figure2}. The main difference to the $(+,-)$ case is that the soft masses are Planck suppressed unless both the spurion and the matter fields are IR localized. Degeneracy of the scalars of the first two generations requires
an IR localization for the first two generations. The third generation would then be again lighter than the first two. This case switches completely the UV  localization with the IR localization for all fields and fits with the holographic realization of the Nelson-Strassler mechanism of generating mass hierarchies \cite{flavor}.
As an aside comment, in all the regions displayed in Fig.~\ref{figure2} the brane localized and KK gravity
multiplet contributions to scalar squared masses have an identical $\epsilon$ dependence but are further suppressed as $(k/M)^3$ and $(k/M)^2$ respectively.

\section{Summary and open questions}
\label{sec:summary}

We have computed the effective action from integrating out KK modes of 5d gauged supergravity at tree level.
In particular, we have calculated the effective Kahler potential of chiral zero modes originating from bulk hypermultiplets. The form of the effective Kahler potential implies tachyonic soft masses for scalars, irrespective of their localization, if the supersymmetry breaking spurion also arises from a bulk field.  This is the case when the spurion is
a matter-like field like in gauge mediation or is a complex structure modulus in a string theory setup.
The tachyonic soft masses go to zero in the sequestered limit.
The results of the present paper strongly suggests that 5d holographic supersymmetric models have constraints in order to avoid
tachyonic contributions to the scalar masses. Moreover, possible positive contributions localized at the fixed points are insufficient to stabilize the vacuum as they are naturally suppressed with respect to the bulk contributions. Then, radion or additional vector multiplets (Kahler moduli in string theory) are needed to generate
positive contributions counterbalancing the tachyonic ones that we found. A particularly simple way out is to invoke the existence
of additional abelian bulk vector multiplets with $(+,-)$ or $(-,+)$ boundary conditions. The corresponding contributions are positive times the product of the matter field and spurion charges. Moreover, they are generically enhanced by a factor of $(M/k)^2$ compared to the
gravitational contributions and universal for the case of $(+,-)$ boundary conditions for UV localized matter fields. Since fermion
mass hierarchies via localization in RS setups can be realized by UV localization of the first two generations and IR localization of the Higgs and the
third generation, we are naturally driven towards degenerate first two generations. This case can therefore be considered as a solution to the supersymmetric flavor problem. On the other hand, IR localization implies smaller masses for stronger IR localized matter fields. For localization pattern leading to successful fermion mass hierarchies, scalar masses have therefore an inverted hierarchy spectrum: third generation scalars are lighters than the first and the second generations, which are degenerate. Complete scalar mass universality arises if all matter fields are UV localized, which is possible at the prize of loosing the geometrical  explanation for fermion mass hierarchies.
The other case, of boundary condition $(-,+)$ for the bulk vector fields, is also viable by switching completely the UV with the IR localization for all fields, like in the 5d holographic realization of the Nelson-Strassler flavor hierarchy setup \cite{flavor}.

Some additional comments are in order concerning the general framework we have been using:
\begin{itemize}
\item The $U(1)_R$ symmetry which has been gauged
has direct low-energy consequences, in the extreme case that the IR scale $\epsilon\, M_{P}$ is low and the MSSM Higgs fields are near IR-localized bulk fields. Indeed, the custodial $SO(4)$ symmetry of the MSSM Higgs sector is broken
by the gauging. In fact the quartic Kahler potential computed in this work generates operators of the type $(\epsilon\,M_P)^{-2}(H^{\dagger} D_{\mu} H)^2$.
\footnote{Similar contributions are expected from the KK modes of the hypercharge gauge boson, as is well known. Notice that one cannot gauge the full $SO(4)$ gauge symmetry in 5d as it does not commute with $U(1)_R$. }
As is well known, this generate deviations in the $\rho$ parameter and puts a lower limit on the IR scale. In models with a IR scale in the TeV region the Higgs has therefore to be predominantly elementary. This constraint seems to be common to any possible gauged supergravity theory.
\item The limit of extreme localization of fields is subtle for operators generated by KK gravity multiplet. When one field is sharply localized and the other is not ($c_i \rightarrow
\pm \infty$, $c_j$ fixed), the scalar masses have a well-defined limit. However, notice that the scalar masses (\ref{sc6}) become large in the limit where both (matter and spurion) fields are sharply localized towards the same brane (UV or IR). The reason is that
in this limit the couplings to the SUGRA KK multiplets are large and the field theory approximation breaks down. This problem does not arises for additional bulk vector multiplets.
\item There are other Kahler operators of interest, for example ones of the type $X^{\dagger} X (H_i^{\dagger} H_i)^2$,
which can modify the Higgs potential for very low supersymmetry  breaking scale \cite{dumitru}. Such operators are induced and could also be computed by the techniques discussed in the present paper.
\item In all cases where gravity is essential, like supersymmetry breaking or radion/moduli stabilization, the supergravity truncation to zero modes is inconsistent. The exchange of KK states has to be properly taking into account in such cases. 

\end{itemize}

 Finally, we would like to point out that similar effective operators and the implications of the resulting scalar masses were
 studied recently in a local F-theory setup in \cite{eranpablo}. The difference compared to the present work is that the massive states    considered and integrated out in \cite{eranpablo} are charged under the Standard Model and localized by the F-theory fluxes. It would
 be very interesting to consider a global F-theory construction and to integrate out the gravity and eventual abelian vector multiplets,  as in our present paper.


\subsection*{Acknowledgments}

We thank Guillaume Bossard, Pablo Camara and Boris Pioline for useful discussions. The work presented was supported in part by the European ERC Advanced Grant 226371 MassTeV,
by the grants ANR-05-BLAN-0079-02 and the PITN contract PITN-GA-2009-237920.

\vspace*{0.7cm}

\appendix


\section{Wave functions and propagators}

\label{wfprop}

In this appendix we compute the propagators appearing in the calculation of the effective action.
We define the full momentum dependent propagators as
\be
G_{s}(z,z';p^2)\equiv\sum_n\,\frac{f^n_{s}(z)f^n_{s}(z')}{p^2+m_n^2}
\ee
where the sum is over all KK modes (including the zero mode).
The wave functions obey the following equations of motion
\be
\partial_z\, \left(z^{1-2s}\,\partial_z\,f_{s}^{n}\right)+z^{1-2s}\,m_n^2\,f_{s}^n=0\,,
\label{eom}
\ee
as well as orthonormality and completeness relations
\be
\int z^{1-2s}\,f_{s}^nf_{s}^m=\delta^{mn}\,,\qquad z^{1-2s}\,\sum_n f_{s}^n(z) f_s^n(z')=\delta(z-z')\,.
\ee
The boundary conditions are Neumann, $\partial_z\,f_{s}(z_i)=0$, for $s=0,2$ and Dirichlet, $f_{s}(z_i)=0$ for $s=1$. This covers all bosonic fields in the supergravity multiplet: $\tilde h_{\mu\nu}$ and $\chi$ ($s=2$), $B_\mu$ and $A_\mu$ ($s=1$), and $\phi$ ($s=0$).
Combining the wave equations with the completeness relation, we derive the
 equations of motion for the propagators:
\be
\partial_z\left(z^{1-2s}\ \partial_z\, G_{s}(z,z';p^2)\ \right)
-p^2\,z^{(1-2s)}\, G_{s}(z,z';p^2)=-\delta(z-z')
\ee
  The boundary conditions are the same as for the wave functions, in addition one has to impose continuity at $z=z'$, as well as the jump condition
\be
\partial_z\, G_{s}(z,z';p^2)|_{z=z'+\epsilon}-\partial_z\,G_{s}(z,z';p^2)|_{z=z'-\epsilon}=-z'^{\,2s-1}
\label{jump}
\ee
which is obtained by integrating around a small interval at $z=z'$.
It can easily be verified that the solutions are
\be
G_{s}(z,z';p^2)=\frac{z_<^s\, B_s(z_0,z_<)\ z_>^s\,B_s(z_1,z_>)}{B_1(z_0,z_1)}
\label{solutionprop}
\ee
where $z_<$ ($z_>$) is the smaller (larger) of the pair $z,z'$ and the functions $B_s(z_i,z)$ are defined as
\be
B_s(z_i,z)\equiv \frac{\pi}{2}\,\bigl(
	Y_1(q\,z_i)J_s(q\,z)-J_1(q\,z_i)Y_s(q\,z)\bigr)\,,\qquad q=\sqrt{-p^2}\, ,
\ee
where $J_{s}$ and $Y_{s}$ are Bessel functions. Note the property $B_1(z_0,z_1)=-B_1(z_1,z_0)$
as well as the relations
\be
B_2(z,z)=-B_0(z,z)=(q\,z)^{-1}\,,
\label{identity1}
\ee
\bea
B_1(z_0,z)B_2(z_1,z)-B_1(z_1,z)B_2(z_0,z)&=&-(q\, z)^{-1}B_1(z_0,z_1)\,,\nn\\
B_1(z_0,z)B_0(z_1,z)-B_1(z_1,z)B_0(z_0,z)&=&(q\, z)^{-1}B_1(z_0,z_1)\, ,
\label{identity2}
\eea
which follow from the properties of the Bessel functions.

The spectrum can be read off from the poles of $G_{s}$. It is given by the zeroes (in $q$) of $B_{1}(z_0,z_1)$, which for $z_0\ll z_1$ coincide to very good approximation with the zeroes of $J_1(q\,z_1)$.
The spectrum of the heavy KK modes is identical for all $s$, as expected from the bulk $N=2$ supersymmetry. In addition, for $s=0,2$, the numerator provides the poles at $q=0$ that correspond to the zero modes of the graviton and the radion. No such pole is present for $s=1$.

Let us note the following relations that will be of use later on,
\bea
\partial_z G_0(z,z_1;p^2)&=&z_1^{-1}\, \frac{B_1(z_0,z)}{B_1(z_0,z_1)} \ , \nn \\
\partial_z G_2(z,z_1;p^2)&=&z^2\,z_1\, \frac{B_1(z_0,z)}{B_1(z_0,z_1)} \ ,
\label{one}
\eea
as well as
\bea
G_0(z_1,z_1;p^2)&=&-\frac{1}{z_1\,q}\,\frac{B_0(z_0,z_1)}{B_1(z_0,z_1)}\nn\\
G_2(z_1,z_1;p^2)&=&\frac{z_1^3}{q}\,\frac{B_2(z_0,z_1)}{B_1(z_0,z_1)} \ ,
\label{two}
\eea
which can be easily checked using the explicit form of the $G_s$, Eq.~(\ref{solutionprop}), as well as the relation Eq.~(\ref{identity1}).
Finally, we notice
\bea
\partial_z\partial_{z'}G_2(z,z')&=&z^3\delta(z-z')-p^2\,z\,z'\,G_1(z,z';p^2) \ , \nn\\
\partial_z\partial_{z'}G_0(z,z')&=&z^{-1}\delta(z-z')-\frac{p^2}{z\,z'}\,G_1(z,z';p^2) \ .
\label{three}
\eea
The equalities for $z\neq z'$ follow straightforwardly from the explicit form of $G_s$.
In order to see the delta functions, we integrate over an infinitesimal interval
\begin{multline}
\int_{z-\epsilon}^{z+\epsilon} dz'\ \partial_{z'}\partial_{z}\,G_{s}(z,z';p^2)\\
=\partial_z\, G_s(z,z';p^2)|_{z'=z+\epsilon}-\partial_z\,G_s(z,z';p^2)|_{z'=z-\epsilon}
 = z^{2s-1} \ ,
\end{multline}
where the last equality follows from Eq.~(\ref{jump}).

\section{Evaluation of the effective action}
\label{zeromomentum}

In this appendix we present details on the evaluation of the effective action, Eq.~(\ref{Leffgraviton}) and Eq.~(\ref{Leffgraviphoton}) which we split according to
\bea
\mathcal L^{\rm graviton}_{\rm eff}&=&\mathcal L^{\tilde h_{\mu\nu}}_{\rm eff}+\mathcal L^{\chi}_{\rm eff}+\mathcal L^{B_\mu}_{\rm eff}+\mathcal L^{\phi}_{\rm eff} \ , \nn\\
\mathcal L^{\rm graviphoton}_{\rm eff}&=&\mathcal L^{A_\mu}_{\rm eff}+\mathcal L^{A_5}_{\rm eff} \ .
\eea
It will prove convenient to define the following, integrated versions of the different components of the 5D energy momentum tensor appearing in Eq.~(\ref{Leffgraviton}).
\bea
\Theta_{\mu\nu}(x,z)&=&\int_{z_0}^z dz'\, {z'}^{-3}\, \tilde T_{\mu\nu}(x,z') \ , \nn\\
\Theta_{tr}(x,z)&=&\int_{z_0}^zdz'\ {z'}^{-3}\, T_{\rho\rho}(x,z')\, , \nn \\
\Theta_{\mu 5}(x,z)&=&\int_{z_0}^z dz' \,{z'}^{-2}\,T_{\mu5}(x,z')\, , \nn \\
\Theta_{55}(x,z)&=&\int_{z_0}^zdz'\, z'^{-1}\tilde T_{55}(x,z')\, .
\eea
In particular, the quantities $\Theta_{\mu\nu}(x,z_1)$, $\Theta_{tr}(x,z_1)$
 and $\Theta_{55}(x,z_1)$ are the 4D operators coupling to the zero modes of $\tilde h_{\mu\nu}$, $\chi$ and $\phi$ respectively.

Following the procedure used in Ref.~\cite{Cabrer:2011fb}, we can integrate by parts in Eq.~(\ref{Leffgraviton}),
\begin{multline}
\int_{z_0}^{z_1} dz\,dz'\ \Theta'(z)\,\Theta'(z')\ G_s(z,z';p^2)
= \int_{z_0}^{z_1} dz\,dz'\ \Theta(z)\,\Theta(z')\ \partial_{z'}\,\partial_z\, G_s(z,z';p^2)\\
-2\,\Theta(z_1)\int_{z_0}^{z_1}dz\ \Theta(z)\ \partial_z\,G_s(z,z_1;p^2)+\Theta(z_1)^2\ G_s(z_1,z_1;p^2) \ ,
\label{ibp}
\end{multline}
where we have used that by definition $\Theta(z_0)=0$.

Let us start with the case $s=1$, i.e.~the effective action resulting from integration of the KK modes of the fields $B_\mu$. The second row in Eq.~(\ref{ibp}) vanishes for $s=1$ since $B_1(z_1,z_1)=0$ (the field $B_\mu$ has Dirichlet boundary conditions). Then
\bea
G_1(z,z';p^2)&=& \frac{(z_<^2-z_0^2)(z_1^2-z_>^2)}{2(z_1^2-z_0^2)}
	+\mathcal O(p^2) \ , \label{G1expansion}\\
\partial_z\,\partial_{z'}\, G_1(z,z';p^2)&=&
z\,\delta(z-z')-\frac{2\,z\,z'}{z_1^2-z_0^2}+\mathcal O(p^2)
\label{ddG1}
\eea
and hence from Eq.~(\ref{Leffgraviton}) one obtains
\be
\mathcal L_{\rm eff}^{B_\mu}=\frac{1}{2\,M^3} \left(\int_{z_0}^{z_1} dz\ z\,\left[\Theta_{\mu 5}(z)\right]^2           			
	-\frac{2}{z_1^2-z_0^2}\left[\int_{z_0}^{z_1} dz\ z\,\Theta_{\mu 5}(z)
			\right]^2\right)+\dots
\ee
where the ellipsis denotes terms with $\partial_\mu$ derivatives acting on $\Theta_{\nu 5}$ which will not be needed for the present work.
For $s=0,2$, we use Eq.~(\ref{one}), (\ref{two}) and (\ref{three}) in Eq.~(\ref{ibp}). The momentum expansions are ($s=0,2$)
\bea
\partial_z\,G_s(z,z_1;p^2)&=&z^{2s-1}\,\Omega_{2-2s}(z)+\mathcal O(p^2) \ , \\
G_s(z_1,z_1;p^2)&=&\frac{2}{(z_1^{2-2s}-z_0^{2-2s})\,p^2}+\int_{z_0}^{z_1}dz\ z^{2s-1}\,\Omega_{2-2s}^2+\mathcal O(p^2)
 \ , \eea
where the functions $\Omega_\alpha(z)$ where defined in Eq.~(\ref{Omegadef}). The leading terms in the last row are the poles the result from the zero modes present in the propagators. These terms should be subtracted.
The final result can thus be written as:
\bea
\mathcal L^{\tilde h_{\mu\nu}}_{\rm eff}&=&\frac{1}{4\,M^3}\int_{z_0}^{z_1} dz\ z^3\,
   \left[\Theta_{\mu\nu}(z)-\Omega_{-2}(z)\,\Theta_{\mu\nu}(z_1)\right]^2+\dots\nn\\
\mathcal L^{ \chi}_{\rm eff}   &=&  -\frac{1}{16\, M^3}\int_{z_0}^{z_1} dz\ z^3\,
        \left[\Theta_{tr}(z)-\Omega_{-2}(z)\,\Theta_{tr}(z_1)\right]^2+\dots\nn\\
\mathcal L^{\phi}_{\rm eff}   &=&  \frac{1}{6\, M^3}\int_{z_0}^{z_1} dz\ z^{-1}\,
        \left[\Theta_{55}(z)-\Omega_2(z)\,\Theta_{55}(z_1)\right]^2+\dots
        \label{final}
\eea
where again the dots denote terms with 4d derivatives acting on the $\Theta'$s.

In full analogy we can write the results for the integration of the graviphoton. This was already given in Ref.~\cite{Cabrer:2011fb}. In terms of the quantities
\bea
\mathcal J_\mu(x,z)&=&\int_{z_0}^z dz'
z'^{-3}J_\mu(x,z')\,,\nn\\
\mathcal J_5(x,z)&=&\int_{z_0}^z dz'
z'^{-2}J_5(x,z')\,,
\label{defcur}
\eea
it reads
\bea
\mathcal L^{ A_\mu}_{\rm eff}&=&
  \frac{1}{2}\int_{z_0}^{z_1} dz\ z\,\bigl[\,\mathcal J_{\mu }(z)\,\bigr]^2
           			-\frac{1}{z_1^2-z_0^2}\left[\int_{z_0}^{z_1} dz\ z\,\mathcal J_{\mu}(z)
			\right]^2+\dots\nn\\
\mathcal L^{A_5}_{\rm eff}   &=&  \frac{1}{2}\int_{z_0}^{z_1} dz\ z^{-1}\,
        \bigl[\,\mathcal J_5(z)-\Omega_2(z)\,\mathcal J_5(z_1)\,\bigr]^2+\dots
       \label{Leffgraviphoton2}
\eea
%


\section{Effective action for scalar zero modes}

\label{app:scalars}

First of all we display here evaluate some integrals necessary in order to parameterize the result.
We define the (normalized) wave function
\be
f_i (z) = \sqrt{\frac{1-2c_i}{z_1^{1-2c_i}-z_0^{1-2c_i}}}\ z^{\frac{3}{2}-c_i}
\label{scalar1}
\ee
and the associated "kinetic distribution"
\be
\omega_i(z)\ = \ z^{-3} \ f_i^2(z) \ . \label{scalar2}
\ee
Wavefunction normalization implies that the integrated kinetic distribution
\be
\Omega_{1-2c_i} (z) \ = \ \int_{z_0}^z dw\ \omega_i (w)
\label{scalar3}
\ee
satisfies $\Omega_{1-2c_i}(z_1)=1$.

Let us define $\alpha_{ij}$ as
\be
\alpha_{ij}\equiv
\frac{(3-2c_i)(3-2c_j)}{4\,M^3}\left[\int_{z_0}^{z_1} z\,\Omega_{1-2c_i}\Omega_{1-2c_j}
	-\frac{2}{z_1^2-z_0^2}\int_{z_0}^{z_1} z\, \Omega_{1-2c_i}\int_{z_0}^{z_1} z\, \Omega_{1-2c_j} \right] \ .
	\label{defalpha}
\ee
By explicit calculation we obtain
\be
\alpha_{ij}=\frac{1}{4\,M_{P}^2\epsilon^2}
\frac{(1-2c_i)(1-2c_j)}{(4-2c_i-2c_j)}\frac{(1-\epsilon^{3-2c_i})(1-\epsilon^{3-2c_j})}{(1-\epsilon^{1-2c_i})(1-\epsilon^{1-2c_j})}
-\frac{1}{4\,M_{P}^2}\frac{(3-2c_i)(3-2c_j)}{(4-2c_i-2c_j)} \ , \label{alpha}
\ee
where we have used Eq.~(\ref{RSPlanck}).
A relation that will be important is the following
\be
\int z^{-3}f_i^2f_j^2=  4\, M^3\, \alpha_{ij}+
\frac{2}{1-\epsilon^2} \ ,
	\label{relation}
\ee
which can be checked by explicitly evaluating the integral.

%
\subsection{Dimension-six scalar operators}

 The starting point is the bosonic Lagrangian
\bea
\mathcal L_S^{\Sigma}&=&-\frac{z^{-3}}{1-\frac{\kappa}{2}\,|\phi|^2 }\biggl(
		|D_M\,\phi_i|^2
		+\frac{\kappa}{2}\,\frac{|\phi_i^\dagger\,D_M\phi_i|^2}{1-\frac{\kappa}{2}\,|\phi|^2}
		\biggr)  \nonumber \\
&& -z^{-5}\,m_i^2|\phi_i|^2+m_i^{(1)}\,z^{-4}_1|\phi_i|^2\delta(z-z_1)
			-m_i^{(0)}\,z^{-4}_0|\phi_i|^2 \delta(z-z_0) + \cdots\, ,
\eea
where $|\phi|^2 = \sum_i |\phi_i|^2$ and the bulk and brane masses are given as
\be
m_i^2 = - \left(\frac{3}{2}-c_i\right)\left(\frac{5}{2}+c_i\right)\,,\qquad m_i^{(\alpha)}=\left(\frac{3}{2}-c_i\right)
\ , \ee
whereas $\cdots$ are higher-order terms which are not needed for our purposes.
In order to compute the KK exchange, we need only the covariant quadratic Lagrangian for complex scalars.
The sources are then computed to be
\bea
T_{MN}&=& -\eta_{MN}\left[|D \phi_i|^2+z^{-2}\,m_i^2|\phi_i|^2 \right]
				+D_M \phi^\dagger_i D_N \phi_i
				+D_N \phi_i^\dagger D_M \phi_i \nn\\
			&&+ \delta_M^{\mu} \delta_N^{\nu} \eta_{\mu\nu}\left[m_i^{(1)}\,z^{-1}_1|\phi_i|^2\delta(z-z_1)
			-m_i^{(0)} \,z^{-1}_0|\phi_i|^2\delta(z-z_0)\right]
				\, , \nn \\
J_M&=& i \left( \phi_i^\dagger D_M \phi_i-D_M \phi_i ^\dagger \phi_i \right) \ .
\eea
One obtains ($z<z_1$)
\bea
\Theta_{\mu\nu}(x,z)&=&
			\Omega_{1-2c_i}(z)\ \left[
				D_{\mu} \phi_i^\dagger(x) D_{\nu}\phi_i(x)+ D_{\nu}\phi_i^\dagger(x) D_{\mu}\phi_i(x)
				-\frac{1}{2}\ \eta_{\mu\nu}\ D_{\rho} \phi_i^\dagger D_{\rho} \phi_i + \dots
			 \right] \ , \nn\\
\Theta_{tr}(x,z)&=& 
			-2\ \Omega_{1-2c_i} (z)\ 
				|D_{\mu} \phi_i(x)|^2
			-2\,(3-2c_i)\ z^{-4}\ [f_i(z)]^2\ |\phi_i(x)|^2
			+\mathcal O(\phi^4) \ , \nn\\
\Theta_{55}(x,z)&=& 
				(3-2c_i)\ z^{-2}\  [f_i(z)]^2\ |\phi_i(x)|^2 +\mathcal O(\phi^4) \ ,
			\nn\\
\Theta_{\mu 5}(x,z)&=&
				\frac{1}{2}\,(3-2c_i)\ \Omega_{1-2c_i}(z)\ \partial_\mu|\phi_i(x)|^2 \ .
\eea
Due to the additional contribution from $T_{\mu\nu}$ at $z=z_1$ we get some cancellations:
\be
\Theta_{tr}(x,z_1)=-2\, \mathcal |D_\mu\,\phi(x)|^2\,,\qquad \Theta_{55}(x,z_1)=0\, .
\label{cancellations}
\ee
Since the zero modes of $\chi$ and $\phi$ have constant profiles, Eq.~(\ref{Lag}) implies that they are sourced precisely by
 $\Theta_{tr}(x,z_1)$ and $\Theta_{55}(x,z_1)$ respectively. In particular, the radion does not couple to massless scalar fields at all (at linear order).

For the following, we define the three quartic two-derivative operators
\bea
\mathcal O^1_{ij}&=&|\phi_i|^2|\partial_\mu\phi_j|^2+|\phi_j|^2|\partial_\mu\phi_i|^2 \ , \nn\\
\mathcal O^2_{ij}&=&\phi^\dagger_i\phi_j\partial_\mu\phi_i\partial_\mu\phi_j^\dagger
					+\phi^\dagger_j\phi_i\partial_\mu\phi_j\partial_\mu\phi_i^\dagger \quad , \quad
\mathcal O^3_{ij}\ =\ \partial_\mu|\phi_i|^2\ \partial_\mu|\phi_j|^2 \ . 					
\eea
which are all symmetric under exchange of $i$ and $j$.
\subsubsection*{Direct contribution}

The 5d scalar manifold metric, after truncation, reads
\bea
g_{ij}&=&\frac{\delta_{ij}}{1-\frac{\kappa}{2}|\phi|^2}+\frac{\kappa}{2}\frac{1}{(1-\frac{\kappa}{2}|\phi|^2)^2}\phi_i^\dagger\phi_j\\
	  &\approx&  \delta_{ij} \left(1+\frac{\kappa}{2}\sum_k|\phi_k^2|\right) + \frac{\kappa}{2}\phi^\dagger_i\phi_j \ .
\eea
The direct quartic terms in the action are then given by
\be
\mathcal L^{\rm direct}=
-\frac{1}{4\,M^3}
	\int z^{-3}f_i^2f_j^2
\frac{1}{2}\left(\mathcal O_{ij}^1+ \mathcal O^2_{ij} \right)\\
=-\left[\frac{1}{4\, M_{P}^2}
+\frac{1}{2}\alpha_{ij}
\right](\mathcal O_{ij}^1+ \mathcal O^2_{ij} ) \ .
\ee

\subsubsection*{Contribution from $\chi$}

Since $\Theta_{tr}(z_1)=-2\sum_i\,|D_\mu\,\phi_i(x)|^2$ is already second order in the 4d derivative, in Eq.~(\ref{final}) the term $~\Theta_{tr}(z_1)^2$ does not contribute. The remainder gives
\begin{multline}
\mathcal L^{\chi}_1
=-\frac{1}{4\, M^3}\int z^{-1}\,
\left[(3-2c_j)(\Omega_{1-2c_i}(z)-\Omega_{-2}(z))[f_j(z)]^2
\right]   |D_{\mu}\phi_i(x)|^2|\phi_j(x)|^2 + (i\to j) \ .
\end{multline}
We can rewrite the integral as
\begin{multline}
\int z^{-1}(3-2c_j)f_j^2(\Omega_{1-2c_i}-\Omega_{-2})=
\int [\partial_z f_j^2](\Omega_{1-2c_i}-\Omega_{-2})=
\\
=-\int  f_j^2(\Omega'_{1-2c_i}-\Omega'_{-2})=-
\int \left(f_j^2 f_i^2 z^{-3}-f_j^2\frac{(-2)z^{-3}}{z_1^{-2}-z_0^{-2}}
\right)\\
=-\int z^{-3}\, f_j^2 f_i^2 -2\frac{1}{\epsilon^2-1} = - 4\, M^3 \alpha_{ij} \ .
\end{multline}
Notice that the boundary term in the partial integration vanishes because $\Omega_\alpha(z_0)=0$ and
$\Omega_\alpha(z_1)=1$. In the last equality we have used Eq.~(\ref{relation}).
Importantly, the integral is {\em symmetric} in $i$ and $j$.
We thus can write the result as $\mathcal L_1^{\chi} = \alpha_{ij}\ \mathcal O^1_{ij}$.
There is another contribution from the $p^2$ terms in the propagator. This term has thus some 4d  derivative acting on the sources and is not yet contained in Eq.~(\ref{final}). We have thus to go back to Eqns.~(\ref{Leffgraviton}) and (\ref{ibp}). Since $\Theta_{tr}(z_1)=\mathcal O(\partial_\mu^2)$ already, we can focus on the first line in Eq.~(\ref{ibp}). Using Eq.~(\ref{three})
we find
\be
\mathcal L_2^\chi=-\frac{(3-2c_i)(3-2c_j)}{4\,M^3}
\,|\phi_i(x)|^2\,\partial_\mu^2\,|\phi_j(x)|^2
\int_{z_0}^{z_1} dz\,dz'\ z^{-3}\,z'^{-3}\ f_i^2(z)\,
f_j^2(z')\ G_1(z,z';0)\,.
\ee
Writing $z^{-3}f_i^2=\Omega_{1-2c_i}'(z)$, integrating by parts and using Eq.~(\ref{ddG1}) the integral can be brought to the form Eq.~(\ref{defalpha}), giving $\mathcal L_2^{\chi} = \alpha_{ij}\ \mathcal O^3_{ij}\,$. The full result from $\chi$ exchange is therefore
\be
\mathcal L^{\chi} = \mathcal L^\chi_1+\mathcal L^\chi_2=\alpha_{ij}\ ( \mathcal O^1_{ij} + \mathcal O^3_{ij}) \ .
\ee
\subsubsection*{Contribution from $\phi$}

In this case there is no contribution from the zero-momentum part of the propagators. There is however a contribution from the
$p^2$ terms. Since $\Theta_{55}(z_1)=0$, only the first term in Eq.~(\ref{ibp}) contributes there.
One finds using Eq.~(\ref{three})
\be
\mathcal L^\phi=\frac{(3-2c_i)(3-2c_j)}{6\,M^3}\,|\phi_i(x)|^2\partial_\mu^2\,|\phi_j(x)|^2
\int_{z_0}^{z_1}dz\,dz'\
z^{-3}z'^{-3} f_i^2(z)\,f_j^2(z')\, G_1(z,z';0)
\ee
 which equals
\be
\mathcal L^{\phi} = - \frac{2}{3} \alpha_{ij}\ \mathcal O^3_{ij} \ .
\ee

\subsubsection*{Contribution from $A_\mu$}

The contribution comes entirely from the zero-momentum part of the graviphoton propagator, which equals
\begin{multline}
\mathcal L^{A_\mu}=
\frac{3}{8\,M^3}
\left(1-\frac{2}{3}c_i\right)\left(1-\frac{2}{3}c_j\right)\left[
\int_{z_0}^{z_1} z\ \Omega_{1-2c_i} \Omega_{1-2c_j}\right.\\\left.-\frac{2}{z_1^2-z_0^2}\int_{z_0}^{z_1} z\ \Omega_{1-2c_i}\int_{z_0}^{z_1} z\ \Omega_{1-2c_j}
\right]
 J_\mu^i(x)\ J_\mu^j(x) \ = \
\frac{1}{6}\ \alpha_{ij}\
J_\mu^i(x)\ J_\mu^j(x) \ .
\end{multline}
We obtain
\be
\mathcal L^{A_\mu}=\frac{1}{6}\ \alpha_{ij}\ (2\mathcal O^2_{ij}- \mathcal O^3_{ij}) \ .
\ee
\subsubsection*{Contribution from $B_\mu$}

The contribution comes entirely from the $p^2=0$ part of the $B_\mu$ propagator and is given by
\begin{multline}
\mathcal L^{B_\mu}=
\frac{1}{2\,M^3}\frac{(3-2c_i)(3-2c_j)}{4}\left[
\int_{z_0}^{z_1} z\ \Omega_{1-2c_i} \Omega_{1-2c_j}\right.\\
	\left.-\frac{2}{z_1^2-z_0^2}\int_{z_0}^{z_1} z\ \Omega_{1-2c_i} \int_{z_0}^{z_1} z\ \Omega_{1-2c_j}
\right] \partial_\mu|\phi_i(x)|^2\partial_\mu|\phi_j(x)|^2 \ = \ \frac{1}{2}\ \alpha_{ij}\ \mathcal O^3_{ij} \ .
\end{multline}
The result is therefore
\be
\mathcal L^{B_\mu}=\frac{1}{2}\ \alpha_{ij}\ \mathcal O^3_{ij} \ .
\ee

\subsubsection*{The full result}

Adding all bosonic terms, integrating by parts and using equations of motion\footnote{Using the equations of motion in the quartic terms is equivalent to a field redefinition up to higher order terms. The exact form of this field redefinition can easily be checked to be $\Phi_i\to\Phi_i(1+\frac{2}{3}\sum_j \alpha_{ij}|\Phi_j|^2) $, yielding
$\sum_i|D_\mu\,\Phi_i|^2\to \sum_i|D_\mu\,\Phi_i|^2+\frac{2}{3}\sum_{i,j}\alpha_{ij}(O^1_{ij}+O^3_{ij})+\dots $.} (which gives $\mathcal O^3_{ij}=-\mathcal O^1_{ij}$), we finally find
\be
\mathcal L^{scalar} = - \sum_{ij} \left( \frac{1}{4\, M_{P}^2} + \frac{1}{6} \alpha_{ij}\right) \ (\mathcal O_{ij}^1+ \mathcal O^2_{ij} ) \ .
	\label{scalar10}
\ee
This precisely matches the fermionic contribution in order to produce a consistent supersymmetric Lagrangian.



\end{document}